\documentclass[preprint,superscriptaddress,amsmath,amssymb]{revtex4-2}
\usepackage{dcolumn}
\usepackage{bm}
\usepackage{graphicx}
\usepackage{color}
\usepackage{footnote}
\usepackage{lineno}
\usepackage{amsmath}
\usepackage{comment}
\usepackage{subfigure}
\usepackage{multirow} 
\usepackage{wasysym} 
\usepackage{float} 
\usepackage{siunitx}
\begin{document}
\title{Performance of the Fully-equipped Spin Flip Chopper For Neutron Lifetime Experiment at J-PARC}
\author{K.~Mishima}
\email{kenji.mishima@kek.jp}
\affiliation{Institute of Materials Structure Science, KEK, Tokai, 319-1106, Japan}
\affiliation{J-PARC Center, Tokai, 319-1195, Japan}
\affiliation{SOKENDAI, Shonan Village, Hayama, 240-0193, Japan}

\author{G.~Ichikawa}
\affiliation{Institute of Materials Structure Science, KEK, Tokai, 319-1106, Japan}
\affiliation{J-PARC Center, Tokai, 319-1195, Japan}

\author{Y.~Fuwa}
\affiliation{Japan Atomic Energy Agency, Tokai, 319-1195, Japan}

\author{T.~Hasegawa}
\affiliation{Department of Physics, Nagoya University, Nagoya, 464-8602, Japan}

\author{M.~Hino}
\affiliation{Institute for Integrated Radiation and Nuclear Science, Kyoto University, Osaka, 590-0494, Japan}

\author{R.~Hosokawa}
\affiliation{Research Center for Advanced Particle Physics (RCAPP), Kyushu University, Fukuoka, 819-0395, Japan}

\author{T.~Ino}
\affiliation{Institute of Materials Structure Science, KEK, Tokai, 319-1106, Japan}
\affiliation{J-PARC Center, Tokai, 319-1195, Japan}

\author{Y.~Iwashita}
\affiliation{Institute for Integrated Radiation and Nuclear Science, Kyoto University, Osaka, 590-0494, Japan}

\author{M.~Kitaguchi}
\affiliation{Institute of Materials Structure Science, KEK, Tokai, 319-1106, Japan}
\affiliation{Kobayashi-Maskawa Institute for the Origin of Particles and the Universe (KMI), Nagoya University, Nagoya, 464-8602, Japan}

\author{S.~Matsuzaki}
\affiliation{Department of Physics, Kyushu University, Fukuoka, 819-0395, Japan}

\author{T.~Mogi}
\affiliation{Department of Physics, The University of Tokyo, Tokyo, 113-0033, Japan}

\author{H.~Okabe}
\affiliation{Department of Physics, Nagoya University, Nagoya, 464-8602, Japan}

\author{T.~Oku}
\affiliation{J-PARC Center, Tokai, 319-1195, Japan}
\affiliation{Japan Atomic Energy Agency, Tokai, 319-1195, Japan}
\affiliation{Graduate School of Science and Engineering, Ibaraki University, Mito, 310-8512, Japan}

\author{T.~Okudaira}
\affiliation{Department of Physics, Nagoya University, Nagoya, 464-8602, Japan}
\affiliation{J-PARC Center, Tokai, 319-1195, Japan}
\affiliation{Japan Atomic Energy Agency, Tokai, 319-1195, Japan}

\author{Y.~Seki}
\affiliation{Institute of Multidisciplinary Research for Advanced Materials, Tohoku University, Sendai, 980-8577, Japan}

\author{H.~E.~Shimizu}
\affiliation{SOKENDAI, Shonan Village, Hayama, 240-0193, Japan}

\author{H.~M.~Shimizu}
\affiliation{Department of Physics, Nagoya University, Nagoya, 464-8602, Japan}

\author{S.~Takahashi}
\affiliation{Graduate School of Science and Engineering, Ibaraki University, Mito, 310-8512, Japan}
\affiliation{J-PARC Center, Tokai, 319-1195, Japan}
\affiliation{Japan Atomic Energy Agency, Tokai, 319-1195, Japan}

\author{M.~Tanida}
\affiliation{Department of Physics, Kyushu University, Fukuoka, 819-0395, Japan}

\author{S.~Yamashita}
\affiliation{Research and Regional Cooperation Office, Iwate Prefectural University, Takizawa, 020-0693, Japan}

\author{M.~Yokohashi}
\affiliation{Department of Physics, Nagoya University, Nagoya, 464-8602, Japan}

\author{T.~Yoshioka}
\affiliation{Research Center for Advanced Particle Physics (RCAPP), Kyushu University, Fukuoka, 819-0395, Japan}

\date{\today}
\begin{abstract}
To solve the ``neutron lifetime puzzle,'' where measured neutron lifetimes differ depending on the measurement methods, an experiment with pulsed neutron beam at J-PARC is in progress. In this experiment, neutrons are bunched into 40-cm lengths using a spin flip chopper (SFC), where the statistical sensitivity was limited by the aperture size of the SFC.
The SFC comprises three sets of magnetic supermirrors and two resonant spin flippers. In this paper, we discuss an upgrade to enlarge the apertures of the SFC. 
With this upgrade, the statistics per unit time of the neutron lifetime experiment increased by a factor of 2.8, while maintaining a signal-to-noise ratio of 250-400 comparable to the previous one.

Consequently, the time required to reach a precision of 1~s in the neutron lifetime experiment was reduced from 590 to 170~days, which is a significant reduction in time.
This improvement in statistic will also contribute to the reduction of systematic uncertainties, such as background evaluation, fostering further advancements in the neutron lifetime experiments at J-PARC.

\end{abstract}
%
\maketitle
\section{INTRODUCTION}
\hspace{\parindent}
A neutron transforms into a proton, electron, and antineutrino through $\beta$ decay. The decay lifetime is an essential parameter in cosmology and particle physics. 
For instance, light elements in the universe were formed through big-bang nucleosynthesis (BBN), and the comparison of observations with theoretical predictions of element abundances offers a valuable opportunity to verify cosmological models. The neutron lifetime determines the proton-to-neutron ratio at the onset of the BBN, thereby influencing the yield of light elements, particularly $^4$He~\cite{mathews2005}.
In a standard particle physics model, neutron lifetime is described by the matrix element $V_{\rm ud }$ of the Cabibbo-Kobayashi-Maskawa matrix. Here, $V_{\rm ud }$ can be used to determine the neutron lifetime and the ratio of the axial vector to the vector coupling constants~\cite{PDG2022}. 
Moreover, the neutron lifetime is required for calculating the cross-section of the inverse reaction of neutron $\beta$ decay, which is an antineutrino capture by protons~\cite{mention2011}.

The neutron lifetime is measured using two methods. The first is the beam method, in which the number of incoming neutrons and decays per unit time are counted. The second is the bottle method, wherein neutrons are confined in a container, and their missing over time is observed to derive their lifetime. While the uncertainty of the former measurement is 2~s and that of the latter reaches subseconds, the average values of both methods differ, with a discrepancy of 9.5~s (4.6$\sigma$)~\cite{PDG2022}. Whether this discrepancy results from an error in the experiment or indicates an unknown phenomenon remains unresolved. Thus, experiments that are qualitatively different from previous ones, particularly beam methods that have not yet achieved high precision, are eagerly awaited.

To solve this problem, a different experiment using pulsed neutrons is currently underway at J-PARC~\cite{hirota2020neutron}. This experiment is classified as a beam method. Unlike previous beam experiments that detected protons from neutron decay, this new approach detects both neutrons and electrons from $\beta$ decay using the same detector.
Because the observed particles are different, any oversight in the previous proton measurements should not affect this method.

The experiment is being conducted at the Materials and Life Science Experimental Facility (MLF) of J-PARC on the Fundamental Physics Beamline (BL05/NOP)~\cite{mishima2009,nakajima2017}.
Neutrons are introduced into a 1-m-long gas detector (time projection chamber, TPC) after shaping into bunches with a length of 40~cm so that being fully contained inside the detector. Considering the data only when the neutron bunch is completely inside the TPC, the detector can cover the neutrons with a full $4\pi$ solid angle and achieve high detection efficiency. Moreover, by bunching neutrons, the $\gamma$-ray backgrounds from the interactions of neutrons with the vacuum windows and beam catchers have been reduced~\cite{arimoto2015,hirota2020neutron}.

The experiment utilizes a spin-flip chopper (SFC) to shape the neutrons into bunches. 
An SFC is a device that changes the beam direction using the spin flip of polarized neutrons, and comprises magnetic supermirrors, spin flippers, and a guide coil to maintain polarization.
Among the incident polarized neutron beams, only the neutrons that have not undergone spin flipping are reflected by the magnetic mirrors and transported to the TPC. Consequently, spin-flipped neutrons pass through the mirrors and are removed using neutron absorbers.
By switching the radio-frequency (RF) current applied to the spin flipper at the correct moment, an arbitrary length of bunches can be created.
The SFC changes the neutron beam path by mirror reflections and thus avoids the intense $\gamma$-rays from upstream of the beamline to reach the TPC. This is the reason why we adopt the SFC for the lifetime experiment.
Because SFC uses magnetic mirrors with finite polarization selection, it suffers from the disadvantage of a poor signal-to-noise ratio (S/N) compared with ordinal material choppers. Therefore, in this experiment, using a two-stage combination of magnetic mirrors and flippers, we have achieved an S/N of 300--400~\cite{taketani2011}.

The current precision of the neutron lifetime experiment at J-PARC is 
$\tau_{\rm n}=898~\pm~10_{\rm(stat.)}~^{+15}_{-18}{}_{\rm (sys.)}$
~\cite{hirota2020neutron}.
The neutron intensity obtained by the previous SFC~\cite{taketani2011} was limited by the size of the magnetic mirrors, and it is necessary to improve the statistics to achieve the precision of 1~s, which is the target of this experiment.
In this study, we tripled the number of neutrons introduced into the TPC using larger magnetic supermirrors and flippers of the SFC. The design and performance are discussed in this paper. We expect that the improved performance of SFC will be useful for several other neutron experiments, including measurements of the neutron decay asymmetry~\cite{markisch2019}.

In this paper, after describing the principle of the SFC in Section~\ref{sec:SFCprinciple}, the design of the newly implemented SFC and its elements are discussed in Section~\ref{sec:SFCdesign}, and the results and their evaluation measured with a neutron beam are discussed in Section~\ref{sec:MEASUREMENT}. Section~\ref{sec:TPC} describes the response when the neutron beam was introduced into the TPC using the new SFC.

\section{Operating Principle of the SFC}
\label{sec:SFCprinciple}
In this section, we explain the operating principle of the SFC. For polarized neutrons, a spin flipper can be used to control the direction of neutron using a device with spin-dependent reflectivity (e.g., a magnetic supermirror) to reflect or transmit neutrons by controlling the spin over time.
In contrast to conventional neutron choppers, flippers can be electrically controlled (on and off), thus facilitating high-speed pulse shaping in any form and achieving a fast rise time (with an error function $\sigma$ of \qty{33}{\micro\second})~\cite{taketani2011}.
One challenge is the smaller S/N value, the off-to-on neutron intensity ratio, compared to that of conventional neutron choppers. The S/N of a general neutron chopper that uses shielding materials such as boron exceeds 10$^{6}$~\cite{bewley2011}. 
On the contrary, that of a typical neutron-polarization mirror is approximately 20~\cite{taketani2011}. This limitation can be overcome by using multiple stages of the SFC. Tasaki $\it{et~al.}$ achieved an S/N of 530 for 0.9~nm monochromatic neutrons using two flippers and four magnetic mirrors~\cite{tasaki2003}.

\begin{figure}[ht]
\centering
  \includegraphics[width=10cm]{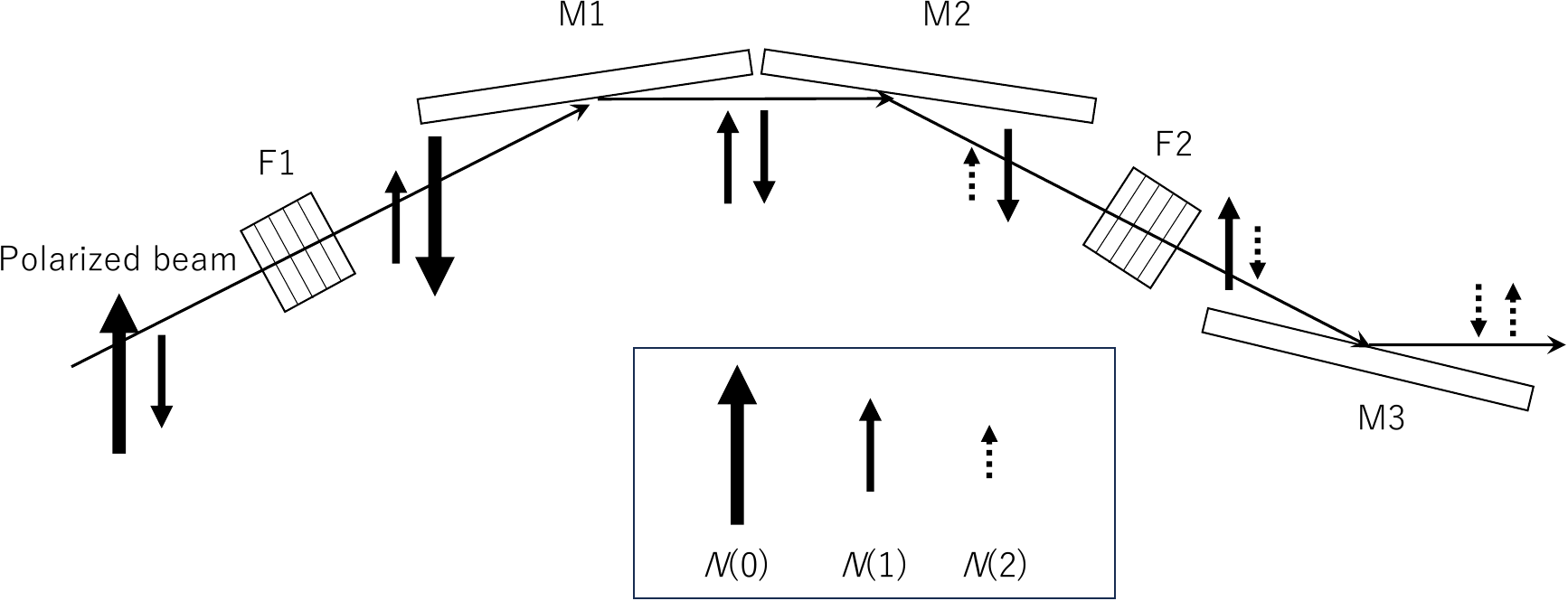}
    \caption{A schematic diagram of the SFC layout. The magnitude scales of $n_{\rm +}$ and $n_{\rm -}$ in the flipper operating mode are indicated by upward and downward arrows, respectively. The number of flipper-mirror pairs passed is shown as $N(i)$.}
  \label{fig:SFCschematic}
\end{figure}

In the previous neutron lifetime experiment, an SFC with two flippers and three magnetic mirrors was used to obtain a practical S/N and flux.
The newly installed  SFC in this study has the same configuration. They were aligned in the order of Flipper~1 (F1), Magnetic Mirror~1 (M1), Magnetic Mirror~2 (M2), Flipper~2 (F2), and Magnetic Mirror~3 (M3). The schematic diagram of the SFC layout is shown in Fig.~\ref{fig:SFCschematic}.

Here, we formulate the number of neutrons transported by the SFC in this setup. 
First, we define the neutron vector \( \vec{n} \) by classifying the neutrons according to their spin components, with \( n_{\rm +} \) representing neutrons whose spins are parallel to the magnetic field and \( n_{\rm -} \) representing those with antiparallel spins, as follows:
\begin{equation}
\vec{n} = 
 \left(
  \begin{array}{c}
     n_{\rm +} \\
     n_{\rm -}
  \end{array}
 \right).
\end{equation}
Then, the neutron polarization $P$ is defined as:
\begin{equation}
P =  \frac{n_{\rm +}-n_{\rm -}}{n_{\rm +}+n_{\rm -}}.
\label{eq:polarization}
\end{equation}
If we denote the spin-flip efficiency of a flipper as $f$, the spin-flip operator $F$ can be described as
\begin{equation}
F =
\begin{pmatrix}
    1-f & f \\
    f & 1-f
\end{pmatrix}.
\label{eq:flip_matrix}
\end{equation}
The reflectivity of the magnetic mirror is expressed as:
\begin{equation}
R = 
\begin{pmatrix}
R_\mathrm{++} & R_{+-}\\
R_{-+} & R_\mathrm{--}
\end{pmatrix},
\label{eq:ref_matrix}
\end{equation}
where $R_{++}$ and $R_{--}$ are the reflectivities of the magnetic mirror for the spin $+$ and spin $-$ components  without spin-flip, respectively, and $R_{+-}$ and $R_{-+}$ indicate the reflections of $+$ ($-$) components with spin-flips, resulting in the $-$ ($+$) state. 
If we define the initial number of neutrons in the SFC as $\vec{n}_{0}$, the number of neutrons in the flipper operating mode (${n}_{\rm on}$) and non-operating mode (${n}_{\rm off}$) are expressed as
\begin{eqnarray}
\vec{n}_{\rm on} = R_3 F_2 R_2 R_1 F_1 \vec{n}_0,  \nonumber \\
\vec{n}_{\rm off} = R_3 R_2 R_1 \vec{n}_0,
\label{eq:SFC}
\end{eqnarray}
where $F_{i}$ and $R_{i}$ denote the $i$th flipper and magnetic mirror, respectively. 
Defining the absolute value of $\vec{n}_{0}$ as the sum of the $+$ and $-$ neutron components from the initial polarization $P_{0}$, we obtain
\begin{equation}
\vec{n}_{0} = 
 \frac{ | \vec{n}_{0} | }{2} 
\begin{pmatrix}
1+P_{0} \\
1-P_{0}
\end{pmatrix}.
\label{eq:nvec0}
\end{equation}

\begin{figure}[ht]
\centering
  \includegraphics[width=10cm]{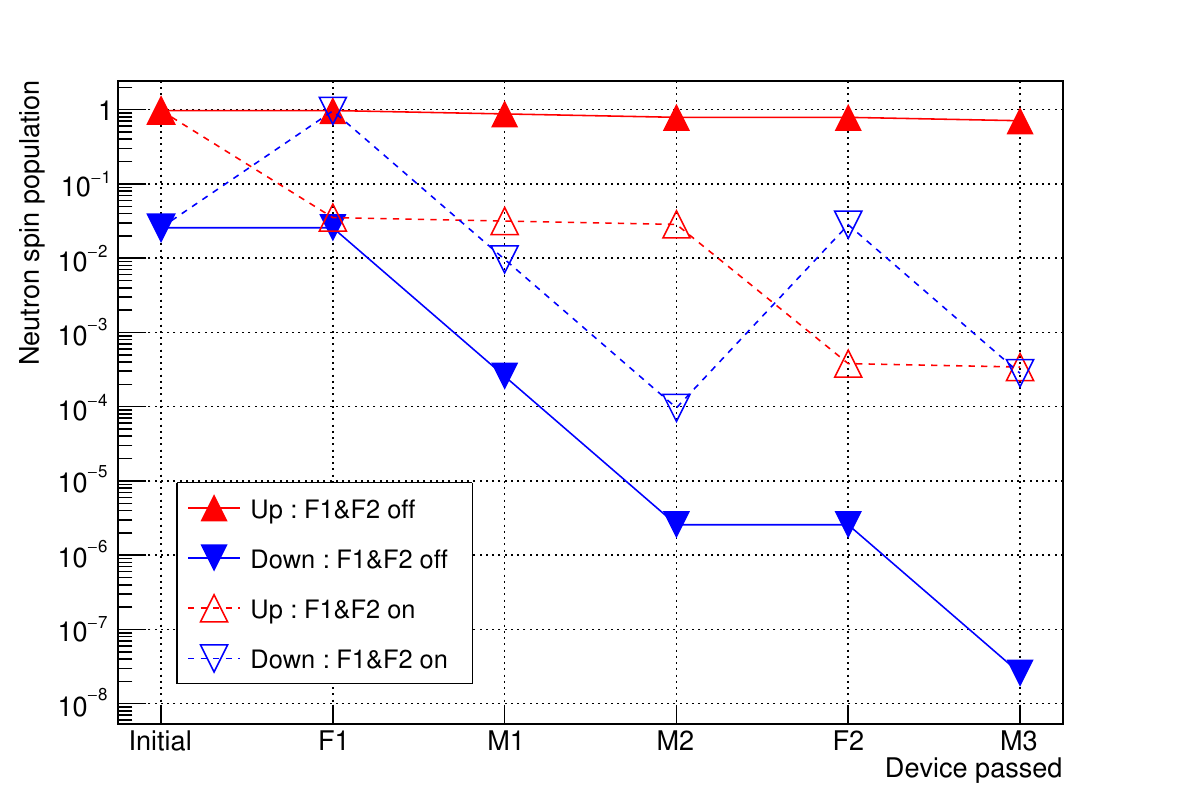}
    \caption{Neutron amounts for each spin component after passing through the device obtained from calculations. The upward (red) and downward (blue) triangles represent $n_{\rm +}$ and $n_{\rm -}$, respectively. The closed triangles with solid lines indicate the case without operating the flippers, while the open triangles with dashed line represents the case where both F1 and F2 are operated.
    }
  \label{fig:M1M2M3}
\end{figure}

Using ideal, but experimentally feasible parameters, 
we set $P_{0}=0.95$, $f=0.99$, $R_{\rm ++}=0.90$, $R_{\rm --}=0.01$, and $R_{\rm +-}=R_{\rm -+}=0$. 
Using Eq.~(\ref{eq:SFC}), we can calculate the neutron intensity after each component of the SFC, as shown in Fig.~\ref{fig:M1M2M3}. For simplicity, it is assumed that $|\vec{n}_{0}|=1$.
Here, the neutron vector after M1 reflection with the flipper-operated, $\vec{n}_{\rm on}(\mathrm{M1})$, is expressed as
\begin{equation}
\vec{n}_{\rm on}(\mathrm{M1}) = R_1 F_1 \vec{n}_0 =  \left( \begin{array}{c} 0.032\\ 0.010 \end{array} \right), \label{eq:nonM1}
\end{equation}
where the $+$ and $-$ components have the same order of magnitude. In this state, the effect of the spin flip is small, and the spin $+$ should be selected again. 
Therefore, M2 is required after M1. 
Thus, ${n}_{\rm on}$ can be reduced by about 1/20 for each pair of flipper and mirrors.
Figure~\ref{fig:SFCschematic} shows how ${n}_{\rm on}$ is reduced for each number of flipper-mirror pairs, denoted as $N(i)$, as indicated by the arrows.
The neutron vector after M3 with flipper operation is expressed as
\begin{equation}
\vec{n}_{\rm on} = \left( \begin{array}{c} 3.4\times10^{-4}\\ 2.8\times10^{-4} \end{array} \right).
\label{eq:nonM3}
\end{equation}
As $|\vec{n}_{\rm off}|$ is 0.71, $\mathrm{S/N} = |\vec{n}_{\rm off}|/|\vec{n}_{\rm on}| = 1140$ can be achieved.

\section{Design of the SFC}
\label{sec:SFCdesign}
This Section describes the overall setup in Subsection~\ref{sec:SFCsetup} and then describes the individual devices in Subsections~\ref{sec:SWPF} and \ref{sec:SpinFlipper}.
The calculation of the spin-flip efficiencies of these devices is presented in subsection~\ref{sec:flipcalc}.

\subsection{Overall Configuration}
\label{sec:SFCsetup}
\begin{figure}[ht]
\centering
  \includegraphics[width=13cm]{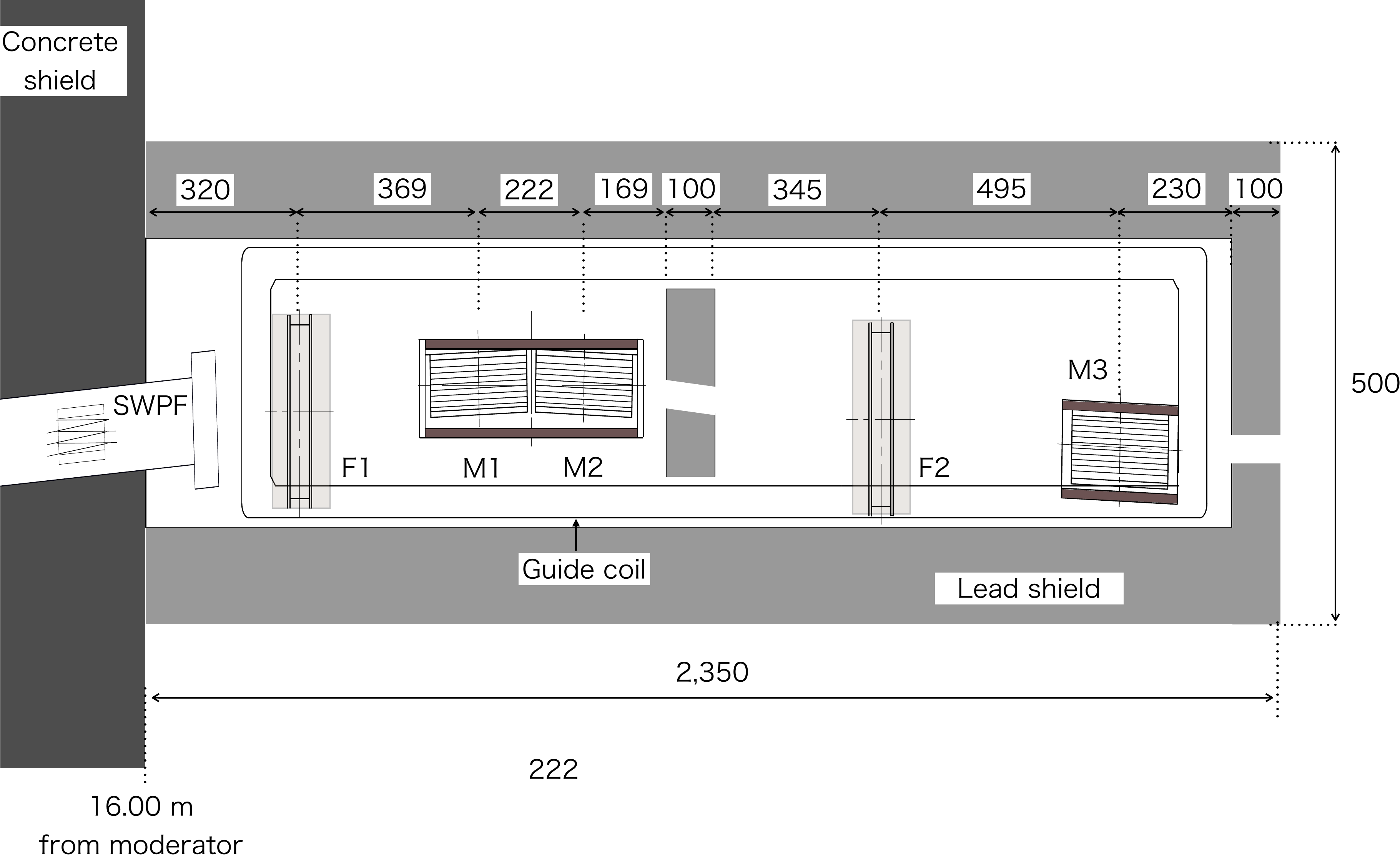}
  \caption{SFC Setup. From upstream, SWPF, flipper 1 (F1), magnetic mirror 1 (M1), magnetic mirror 2 (M2), flipper 2 (F2), magnetic mirror 3 (M3).
 }
  \label{fig:SFCsetup}
\end{figure}

A spallation neutron source at the MLF in J-PARC generates pulsed neutron beams by bombarding a mercury target with 3 GeV protons. The fast neutrons, resulting from spallation reactions, are moderated by liquid hydrogen to cool them to an energy of about 10 meV and are distributed to the neutron beamlines. The neutron lifetime experiment takes place at the BL05 beamline, where the SFC is situated on the polarized branch. The neutron flux at the exit of the beam branch is 
\qty{4.0(0.3)e7}{\per\second\per\cm\squared}
 at a 1~MW equivalent~\cite{nakajima2017}, and the neutron polarization is $P_{0}=94-96\%$~\cite{Ino2011}. In this study, the coordinates in the beam direction are defined as $+Z$, vertically upward as $+Y$, and $X$ as a right-handed system with respect to them. The SFC comprises three sets of magnetic supermirrors and two spin flippers, as shown in Fig.~\ref{fig:SFCsetup}. 
These elements are installed in a 1~mT magnetic field applied by a Helmholtz-type guide coil to maintain neutron polarization. Polarized neutrons have spins parallel or anti-parallel to the $Y$-axis along the guide field.

The SFC is covered with 100-mm-thick lead to shield $\gamma$-rays resulting from neutron capture. The centers of F1 and F2 are located 320 and 1525~mm from the concrete surface (15994~mm from the neutron moderator), and M1, M2, and M3 were positioned at 689, 911, and 2020~mm, respectively.
An additional 100~mm thick lead is installed between M2 and F1. This lead had a beam port of 39.5~mm~$\times$~29.5~mm in vertical and horizontal direction.
The lead downstream of M3 also contained a hole measuring 34.5~mm~$\times$~29.5~mm.
At the entrance and exit of the mirror sets, 5-mm-thick LiF tiles made by sintering a mixture of 95\%-enriched $^6$LiF and PTFE at 30:70~wt\%~\cite{koga2021} are installed.
The neutron beam is collimated to 30~mm~$\times$~35~mm before and after M1M2 and 30~mm~$\times$~40~mm before and after M3. 
The inner walls of the lead shield are covered with rubber containing B$_4$C to absorb the scattered neutrons. The design for neutron transport and $\gamma$-ray shielding was carried out using PHITS~3.20~\cite{sato2023}.

\subsection{Short Wavelength Pass Filter}
\label{sec:SWPF}
\begin{figure}[ht]
\centering
\includegraphics[width=15cm]{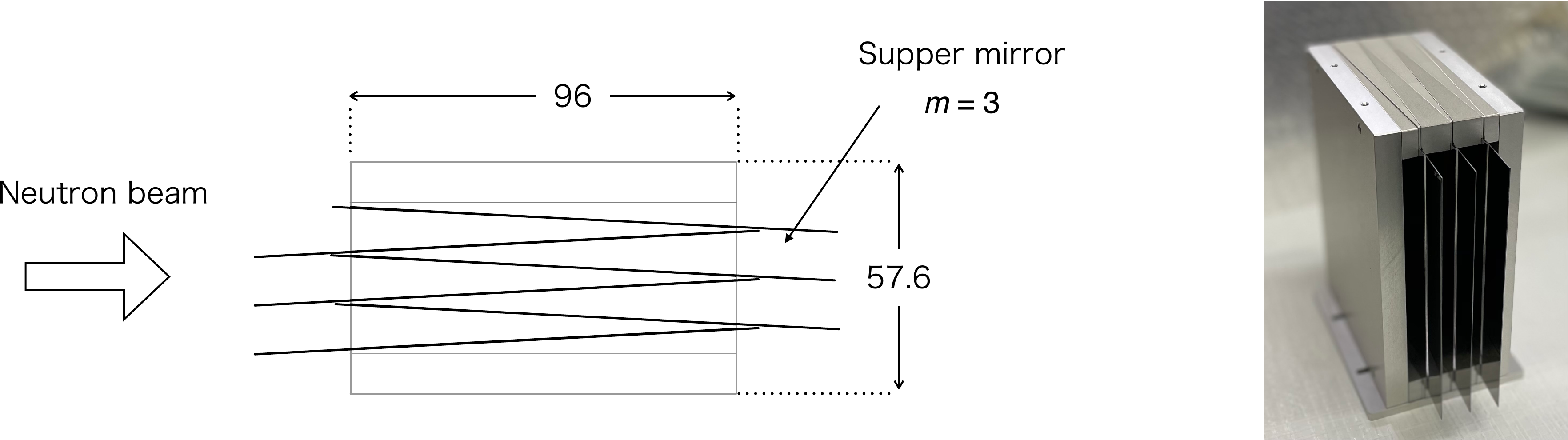}
\caption{Sketch (left) and photo (right) of the short wavelength pass filter.}
  \label{fig:SWPF}
\end{figure}

The pulsed neutron source at J-PARC operates with a repetition rate of 25~Hz.
An event in which neutrons produced by a pulse arrive later than neutrons from the next pulse is referred to as a frame overlap. Because the SFC is designed to flip only neutrons of specific wavelengths within a specific time-of-flight (TOF) interval, it cannot handle neutrons of different wavelengths arriving owing to the frame overlap. 
To remove the slower neutrons arriving as the flame overlapped, a short-wavelength pass filter (SWPF) was installed 20~cm upstream from the exit of the polarized beam branch. The SWPF comprised six supermirrors arranged in a $\Lambda$ shape with an angle of \qty{3}{\degree} relative to the beam, as shown in Fig.~\ref{fig:SWPF}. 
The supermirrors were fabricated on a Si substrate with height, length, and thickness of 110, 125, and 0.3~mm, respectively, using a sputtering apparatus at the Research Reactor Institute, Kyoto University, with an $m$-value (the critical reflection momentum transfer ratio for Ni mirror) of 3~\cite{hino2004}.

\begin{figure}[ht]
\centering
  \includegraphics[width=10cm]{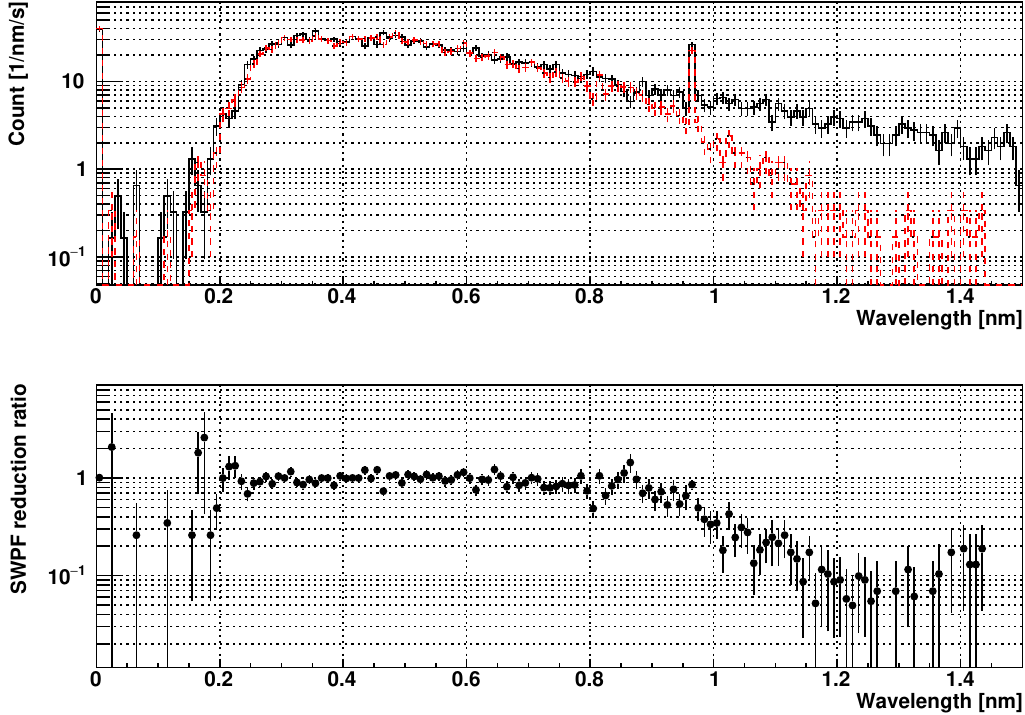}
  \caption{Change in wavelength distribution owing to the short wavelength pass filter (top) and its ratio (bottom). The peak near 1~nm wavelength is owing to the fast neutron component generated when protons collide with the mercury target.}
  \label{fig:SWPFspectrum}
\end{figure}

The performance of the SWPF was evaluated by measuring the counts with and without the filter using a beam monitor (BM) at the exit of the polarized branch (16.4~m). The BM employed was CANBERRA MNH10/4.2F~\cite{ino2014}. The detection efficiency $\varepsilon_{\rm BM}$ followed the so-called $1/v$ law and was $\varepsilon_{\rm BM}=2.63 \times 10^{-4}\cdot(\lambda~\mathrm{[nm]})$ from the measurement compared to a $^3$He proportional counter, where $\lambda$ is neutron wavelength. 
The measured spectra and their ratios are presented in Fig.~\ref{fig:SWPFspectrum}. Wavelengths longer than 1.2~nm were successfully reduced by one order of magnitude.

\subsection{Magnetic Mirror Assembly}
In this upgrade, we used magnetic mirrors manufactured by Swiss Neutronics~\cite{schanzer2016}. These mirrors were created by sputtering an Fe/Si multilayer film onto a Si wafer with 200, 100, and 0.3~mm in length, height, and thickness, respectively. Detailed specifications are listed in Table~\ref{tableMirror}.
\begin{table}[ht]
\centering
\caption{Supermirror Specifications~\cite{schanzer2016}}
\label{tableMirror}
\begin{tabular}{c|c} \hline 
Material & Fe/Si \\
Mirror size $( X, Y, Z )$ [mm] & $( 0.3, 100, 200 )$\\ 
$m$-value & 5.1\\ 
Reflectivity at the edge & 78\% \\
Polarization & 98\% \\ \hline
\end{tabular}
\end{table}

The supermirrors were used as stacks comprising 
8--10 mirrors assembled in the solar-slit style.
For M1 and M2, the mirror configuration comprised eight plates each, while M3 was made of ten plates. The plates were fixed using an aluminum alloy fixture, which were designed to be parallel to the mirrors with a 5.9~mm interval.

\begin{figure}[ht]
\centering
\includegraphics[width=\linewidth]{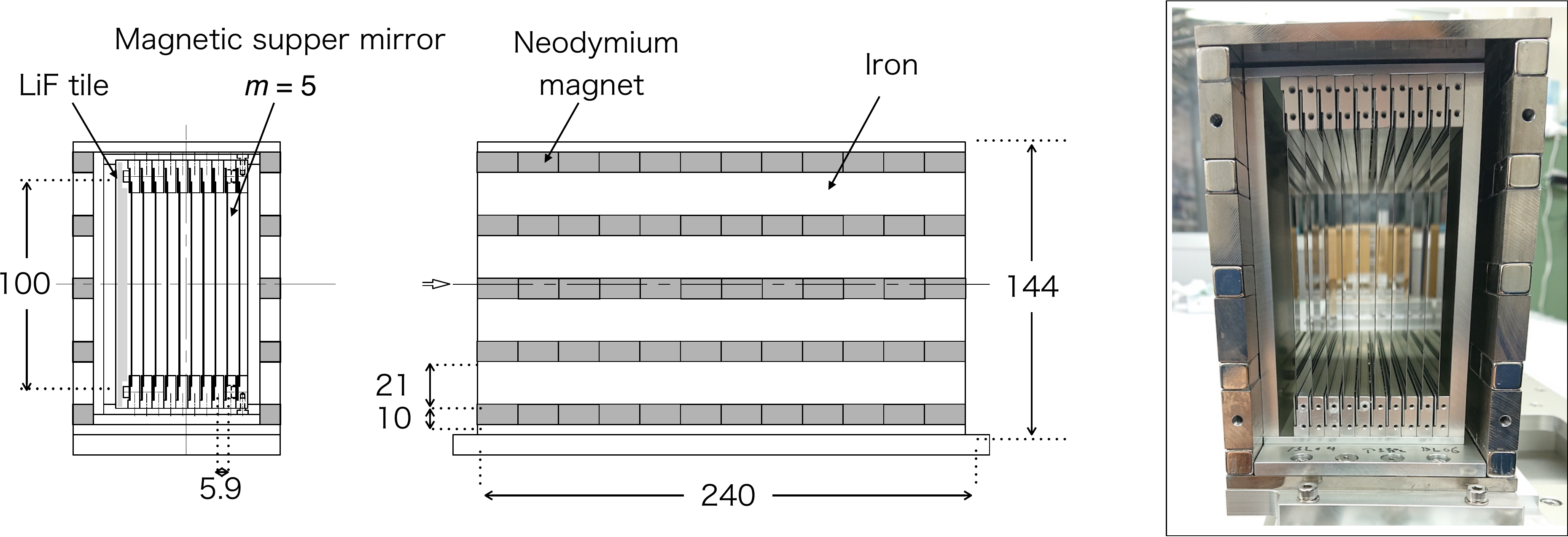}
\caption{Sketch (left, center) and photo (right) of the M3 magnetic mirror holder and mirror stack, viewed from downstream.}
  \label{fig:MirrorHolder}
\end{figure}

To obtain practical reflectivity, magnetic supermirrors must be sufficiently magnetized, which requires a magnetic field of at least 45~mT. Thus, the mirrors were operated in magnetic containers.
In this SFC, adjacent M1 and M2 were housed within the same magnetic container, while M3 was stored separately.
The magnetic container had a structure comprising five neodymium magnets with iron sandwiched between them and capped at the top and bottom (Fig.~\ref{fig:MirrorHolder}). 
The magnetic field measured within the container was 58~mT near the center and 48~mT at the edges for the M1M2 container, and 52~mT at the center and 46~mT at the edges for the M3 container, thereby satisfying the required specification of over 45~mT throughout.

M1, M2, and M3 were located at \qty{-1.6}{\degree},~+\qty{1.6}{\degree}, and \qty{-1.6}{\degree} relative to the polarization beam branch, respectively.
As the polarized branch was tilted \qty{3.2}{\degree} from the beamline axis, the neutrons exiting the SFC were designed to be transported in the same direction as the center axis of BL05 beamline.

\subsection{Spin Flipper}
\label{sec:SpinFlipper}
\begin{figure}[ht]
\centering
  \includegraphics[width=15cm]{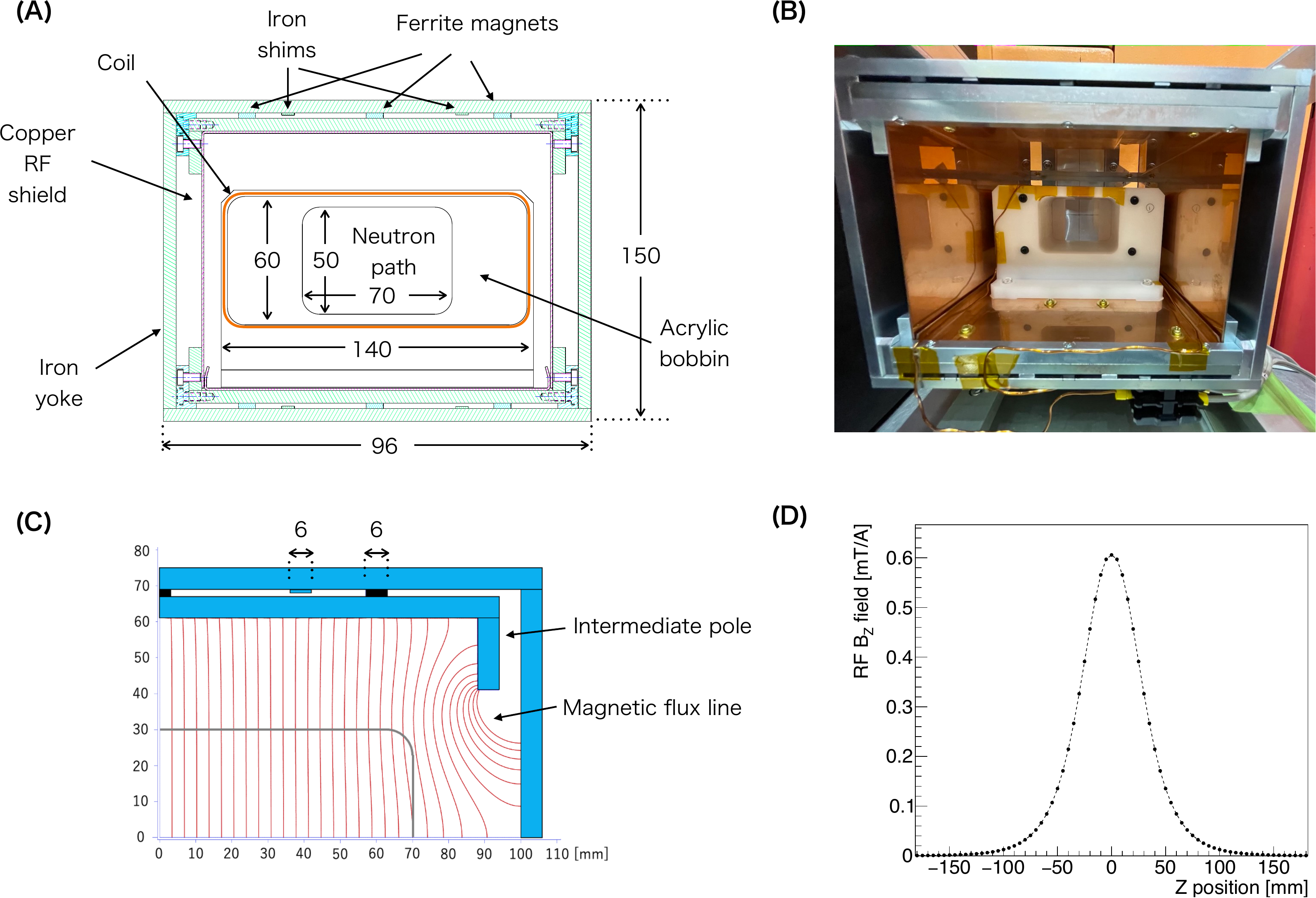}
  \caption{(A) Sketch and (B) photo of a newly developed flipper coil, (C) Diagram of the magnet/shim configuration for applying the flipper $B_{0}$ magnetic field of the flipper and accompanying the simulation results, (D) Simulated RF field strength in the $Z$ direction at the central axis of the coils.}
  \label{fig:Flipper}
\end{figure}

There are two types of neutron spin flippers: adiabatic and resonant. Although the adiabatic type can achieve high flipping efficiency, it is not suitable for the fast switching.
Therefore, we employed a resonant-type spin flipper for the SFC.
In the neutron lifetime experiment, $\gamma$-rays caused by the scattering and absorption of neutrons by wires in the beam axis are a major concern. Therefore, we continued to adopt $Z$-directed air coils as in the previous design. 
In the previous design, a guide coil was used to provide a magnetic field to define the spin quantization axis of the neutrons ($B_{0}$ field)~\cite{taketani2011}. 
In the case of the new SFC, owing to the enlargement of the magnetic containers and the increase in the magnetic field strength, the magnetic field leakage to the coil can no longer be ignored. Thus, they were designed to apply a uniform $B_0$-field (1~mT) by surrounding the flipper coil with an iron magnetic shield and installing ferrite magnets inside a magnetic container. Figures~\ref{fig:Flipper}(A) and (B) show sketches of a newly developed flipper and its downstream view, respectively.
This design is less affected by external magnetic fields, and is compatible with another neutron lifetime experiment using  magnetic fields~\cite{sumi2023}.

Ferrite magnets with residual flux density 300--400~mT, a thickness of 2~mm, and a width of 6~mm were mounted in three rows along the $Z$-axis to generate magnetomotive force. To ensure uniformity, iron poles with thickness and width of 1 and 6~mm, respectively, were mounted. 
The distribution of the magnetic field, determined by a three-dimensional magnetic field simulation, is depicted in Fig.~\ref{fig:Flipper}(C).
The magnetic fields measured in the centers of F1 and F2 with a Gauss meter were 1.18 and 1.08~mT, respectively, 
for a design value of 1.00~mT.
The uniformity of the $B_0$ magnetic field was within 7\% for F1 and 3\% for F2 in the region from the coil center to $X=\pm~20$~mm, $Y=\pm~20$~mm, and $Z=\pm~50$~mm.
A 1-mm thick copper plate was installed around the RF coil, serving as an RF shield and confining the magnetic flux. 

In the previous SFC, coils with diameter and length of 50~mm each were used. To expand the beam size, we enlarged the coil to 60~mm in length and 140~mm in width to apply a uniform RF magnetic field. The fringing field of the RF magnetic field generated by the coil along the $Y$-axis direction was parallel to $B_{0}$, and perpendicular to $B_{0}$ along the $X$-axis direction. The fringing field perpendicular to $B_{0}$ was a significant obstacle to spin-flip efficiency. Therefore, the size in the $X$-direction was increased to avoid this magnetic field disturbance. The calculation of fringing field disturbance is discussed in the next subsection.
The details of the coils are listed in Table~\ref{table:RFcoils}. Figure~\ref{fig:Flipper}(D) presents the RF magnetic field strength in the $Z$-direction at the $XY$ center of the coil, as obtained from the magnetic field simulation.
The RF field intensity in the beam axis was roughly followed a Gaussian distribution, with a standard deviation of 30~mm.

\begin{figure}[ht]
\centering
  \includegraphics[width=10cm]{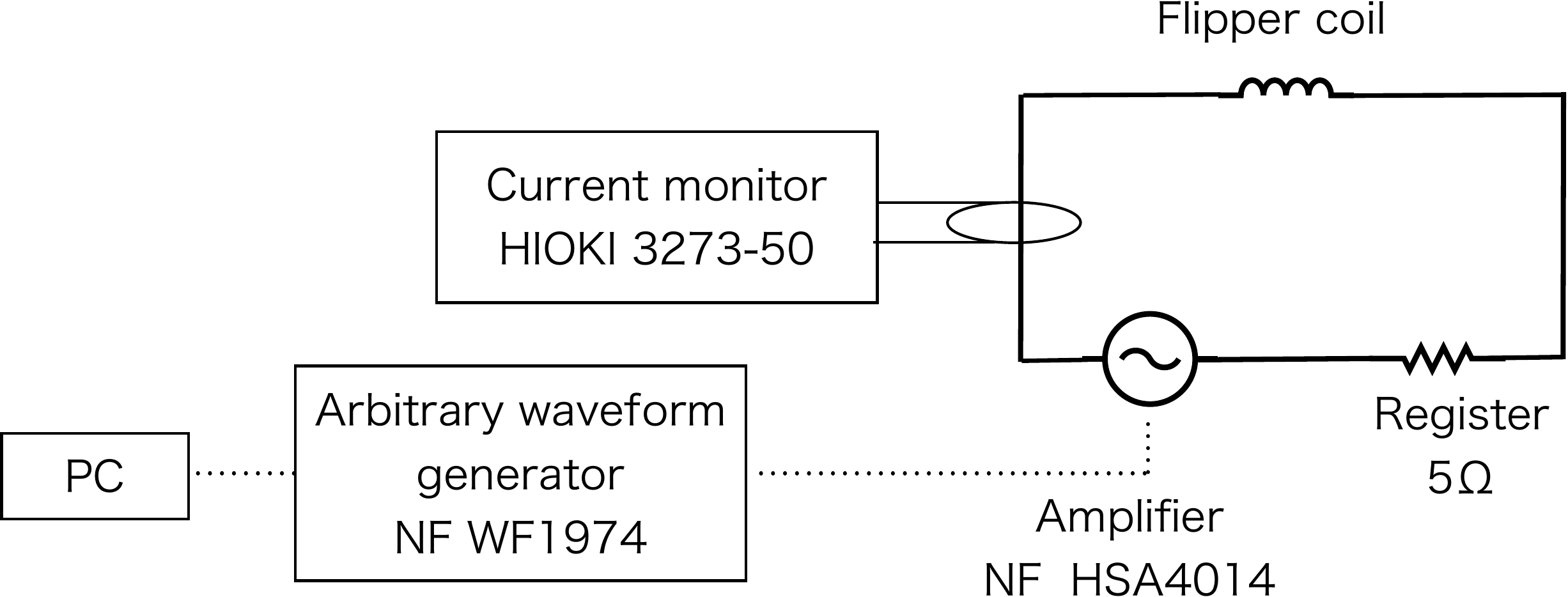} 
 \caption{Schematic diagram of the RF circuit}
  \label{fig:RFcurcuit}
\end{figure}

To operate SFC for pulsed neutrons, the RF current must be controlled according to the TOF. The circuit used to apply a time-dependent RF current is shown in Fig.~\ref{fig:RFcurcuit}.
A waveform was generated using an arbitrary waveform generator, magnified using an amplifier, and applied to the coil. Here, only the resistance of the copper coil wire resulted in a time constant of $\qty{300}{\micro\second}$, which does not satisfy the required flipping time ($\sim$\qty{100}{\micro\second}) for the SFC. Therefore, a 5.0-$\Omega$ resistor was inserted in series in the circuit to shorten the time constant to \qty{20}{\micro\second}.
The frequency corresponding to the applied $B{_0}$ magnetic field was approximately 30~kHz, and the composite impedance at this frequency was 19.5~$\Omega$. As the maximum voltage of the amplifier was $\pm$~50~V, a maximum current of 2.5~A could be applied. 

By optimizing neutron flip efficiency in any region of TOF, the RF current amplitude must be inversely proportional to the neutron velocity.
In addition, the RF current was turned off based on the TOF when creating the bunch. These current controls were realized by inputting an ideal waveform created in 1-\unit{\micro\second} increments into an arbitrary waveform generator.
The equation for the flipping probability is explained in Section~\ref{sec:flipcalc}. 
The actual bunching operation is explained in Section~\ref{sec:bunching_operation}.

\begin{table}[ht]
\centering
\caption{Specifications of the RF coils}
\label{table:RFcoils}
\begin{tabular}{c | c c } \hline
 & F1 & F2 \\ \hline
 Coil size $( X, Y, Z )$ [mm] & \multicolumn{2}{|c}{$( 140, 60, 46 )$}  \\ 
 Diameter of Cu wire [mm] & \multicolumn{2}{|c }{1.2}  \\ 
 Number of turns & \multicolumn{2}{|c}{32}  \\ 
 Inductance in shield (at 1~kHz) [\unit{\micro H}]  & 98.5 & 103.3 \\
 Quality Factor & 2.91 & 2.72 \\
 Coil resistance [$\Omega$] &  0.288  & 0.315 \\
 Additional resistance [$\Omega$] &  5.0  & 5.0  \\
 Phase [degree] &  75.1  & 74.5  \\ \hline
 $B_{0}$ field [mT] &  1.18  & 1.08  \\ \hline
\end{tabular}
\end{table}
 
\subsection{Spin Flip Calculation}
\label{sec:flipcalc}
The spin flipper changes the orientation of the neutron spin by inducing Larmor precession in static and RF magnetic fields. The orientation of the neutron spin $\vec{s}$ in the magnetic field $\vec{B}$ can be described by the following differential equation with respect to time $t$:
\begin{equation}
  \label{eq:lamor}
    \frac{{d\vec{s}}}{{dt}} = \gamma_{\rm n} \vec{s} \times \vec{B},
\end{equation}
where $\gamma_{\mathrm{n}}$ is the neutron gyromagnetic ratio of \qty{1.832e8}{\radian\per\second\per\tesla}~\cite{codata2021} and $\vec{B}$ is the magnetic field vector experienced by the neutrons. 

When the neutron is oriented to the $B_{0}$ field of magnitude $B_{0}$, and an RF magnetic field of angular velocity and amplitude $B_{1}$ perpendicular to it is applied, the spin-flip efficiency $f$ is expressed as the following equation using the applied time of the RF,
$t_{\rm RF}$~\cite{rabi1937,alefeld1981}:
\begin{equation}
  f=\frac{\sin^2 \left(\gamma_{\rm n}B_{1}t_{\rm RF}\sqrt{1+q^{2}}\right)} {1+q^{2}},
  \label{eq:flip_prob}
\end{equation}
\begin{equation}
    q=\frac{2(B_{0}-\omega/\gamma_{\rm n})}{B_{1}}.
  \label{eq:flip_prob2}
\end{equation}
For $q=0$, we have
\begin{equation}
  B_{0}=\omega/\gamma_{\rm n},
  \label{eq:By}
\end{equation}
Consequently, Eq.~(\ref{eq:flip_prob}) can be simplified as
\begin{equation}
  f=\sin^2 \left(\gamma_{\rm n}B_{1}t_{\rm RF}\right),
  \label{eq:sine}
\end{equation}
Thus,
\begin{equation}
  B_{1}=\frac{1}{\gamma_{\rm n}t_{\rm RF}}\left(n+\frac{1}{2}\right)\pi
  \label{eq:Bz}
\end{equation}
This equation yields $f=1$. Here, $n$ is an integer corresponding to the number of times the neutron spin rotates. In this SFC operation, we set $n=0$, implying that it was rotated by $\pi$ within the coil.

\begin{figure}[ht]
\centering
  \includegraphics[width=10cm]{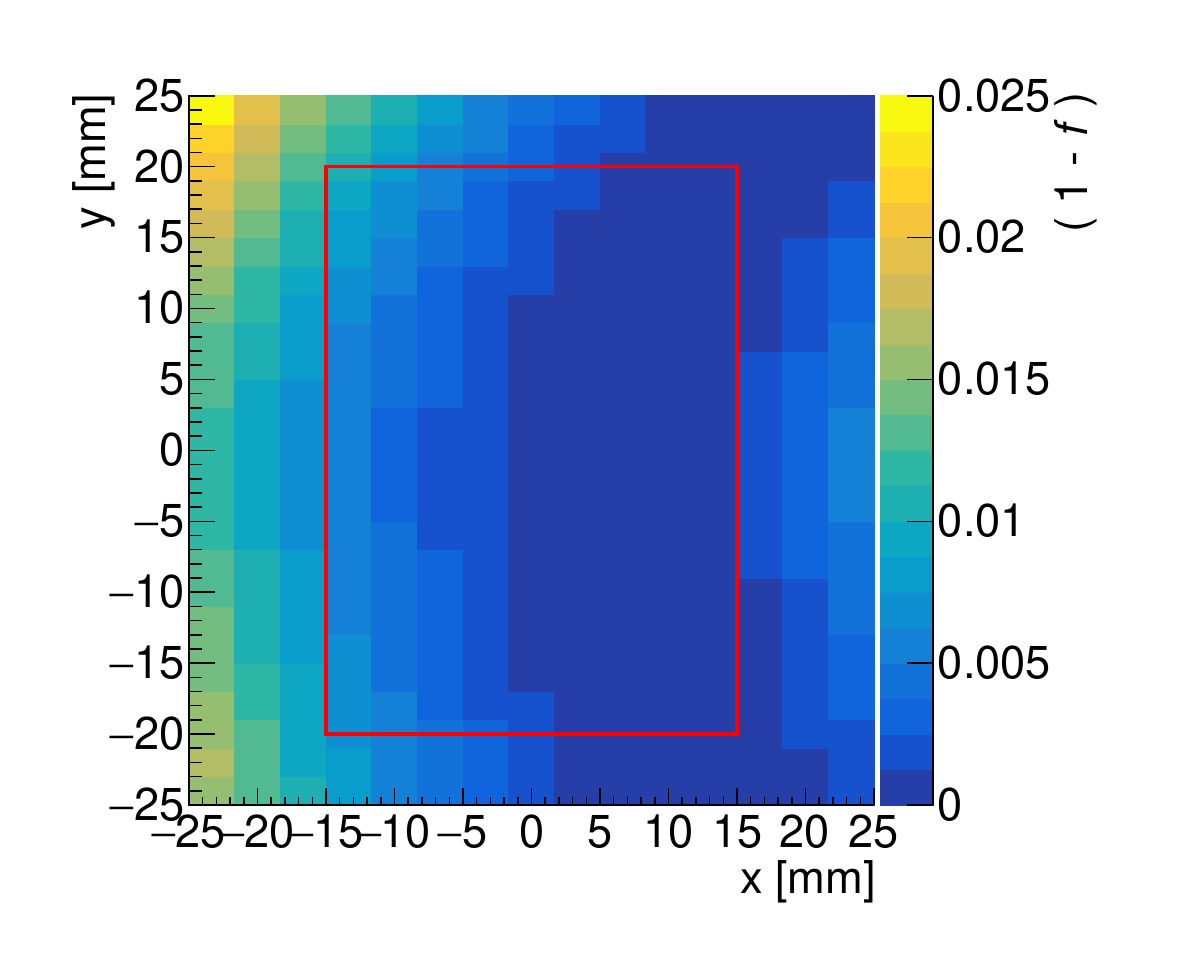} 
 \caption{The results of the flipping efficiency ($1-f$) from the spin flip simulation. The red frame indicates the area of the beam collimator, $30~\mathrm{mm} \times 40~\mathrm{mm}$ in $X$ and $Y$.
 }
  \label{fig:3DSim}
\end{figure}

In reality, these equations are approximations owing to the nonuniformity of the magnetic field or fringing fields from the coil. 
Hence, we derived the spin-flipping efficiency using Eq.~(\ref{eq:lamor}), incorporating the $B_{0}$ and RF field distributions from the magnetic field simulation detailed in Section~\ref{sec:SpinFlipper}. 
The $XY$ distribution of the spin-flip efficiency, represented as $(1-f)$, is shown in Fig.~\ref{fig:3DSim}. In the calculations, the neutron velocity was assumed to be 1,000~m/s, the $B_{0}$ field as 1~mT at the center, and RF frequency as 29.2~kHz. Consequently, the current of the RF coil was optimized to achieve the maximum flipping efficiency.
The results show that for the area within 30~mm~$\times$~40~mm region (red line in Fig.~\ref{fig:3DSim}), where neutrons can pass, the average efficiency was 99.8\% and the worst was 99.0\%.

\section{Measurement}
\label{sec:MEASUREMENT}
\begin{figure}[ht]
 \centering
 \includegraphics[width=14cm]{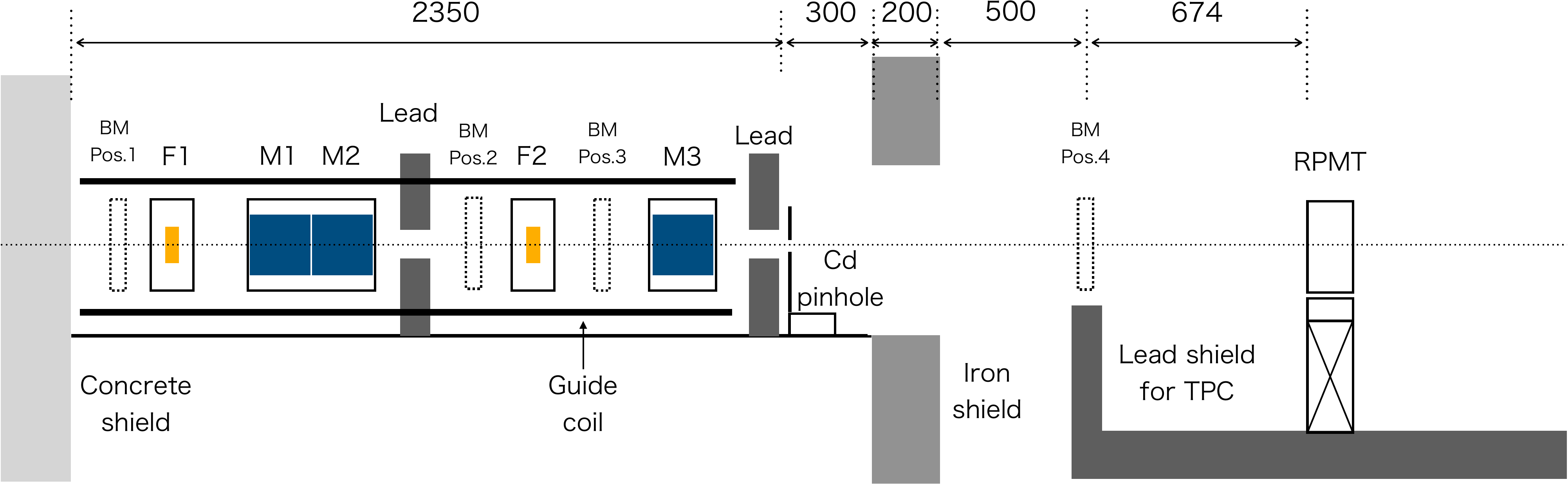}
 \caption{Flux measurement configuration diagram. Measurements were taken while moving the BM to Pos.1-4. The Cd pinhole was only used during the RPMT scan measurement.}
 \label{fig:detector_setup}
\end{figure}

In this Section, the performance of the new SFC is evaluated. The beam fluxes at different positions along the beam path were measured using beam monitor detectors. The position and divergence distribution at the TPC were measured using a two-dimensional detector while scanning the position of the collimator installed at the exit of the SFC lead shield. The flipping efficiencies were optimized by considering the off-to-on ratio of the RF currents.
Finally, beam polarization was measured using a $^3$He spin filter.
The measurement configuration is shown in Fig.~\ref{fig:detector_setup}. The vacuum chamber for the TPC and shielding was removed from these measurements. The results are discussed below.

\subsection{Neutron Flux}
\label{meas_flux}
The neutron flux was measured at four positions: upstream of M1 (Pos.1), downstream of M2 (Pos.2), upstream of M3 (Pos.3), and upstream of the TPC shielding (Pos.4). These results are compared with the beam simulation by PHITS.
The same BM used in Section~\ref{sec:SWPF} was used at Pos.2--4, whereas a different BM with a lower detection efficiency of $8.4\times10^{-6}\cdot(\lambda~\mathrm{[nm]})$ was used at Pos.1 to accommodate a higher count rate. To exclude the spread beam, 
a collimator made of a LiF tile with a $30~\mathrm{mm} \times 40~\mathrm{mm}$ aperture in the $X$ and $Y$ directions was mounted upstream the BM,
and the fluxes were deduced by normalizing the total intensity in this area. The BM counts were converted to neutron flux using $\varepsilon_{BM}$ under the assumption of the $1/v$ law. 
Table ~\ref{table:BMFlux} summarizes the neutron fluxes corresponding to the proton beam power of 1~MW.
Note that the neutron with wavelengths of 0.32--0.78~nm were extracted (which can be reflected by the supermirrors) for comparison between the simulation and measurements.

\begin{table}[ht]
\centering
\caption{Neutron fluxes from the measurements and simulations at each position normalized to a proton beam power of 1~MW, and their ratios.}
\begin{tabular}{p{4.2cm}|p{2.6cm}|p{2.6cm}|p{2.6cm}|p{1.0cm}} \hline 
BM position & Distance from moderator [m] & Measured flux [n/cm$^2$/s] & Simulated flux [n/cm$^2$/s] & Ratio \\ \hline
Pos.1: Upstream M1 & 16.4 & $2.2\times10^{7}$ &  $2.2\times10^{7}$ & 1\\ 
Pos.2: Downstream M2 & 17.3 & $5.9\times10^{6}$ &  $8.0\times10^{6}$ & 0.74\\ 
Pos.3: Upstream M3 & 17.6 & $4.5\times10^{6}$ &  $6.4\times10^{6}$ & 0.71\\ 
Pos.4: Upstream TPC& 19.3 & $1.3\times10^{6}$ &  $1.8\times10^{6}$ & 0.75\\ \hline
\end{tabular}
\label{table:BMFlux}
\end{table}

\begin{figure}[ht]
 \centering
 \includegraphics[width=10cm]{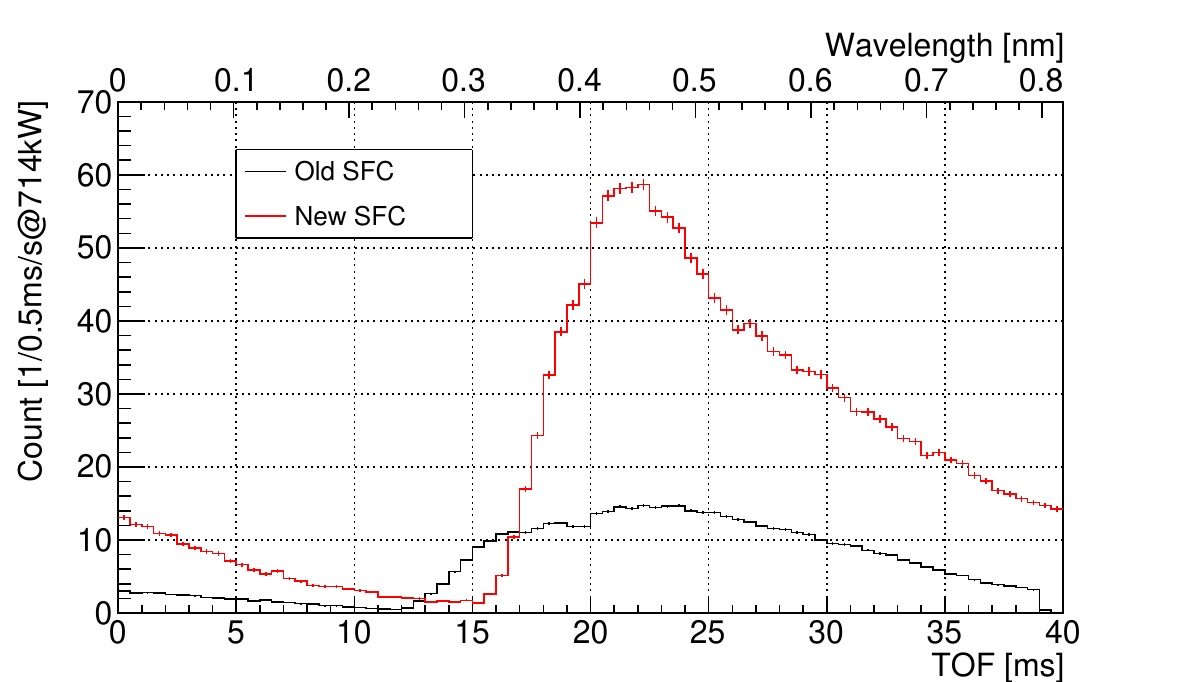}
\caption{Beam intensity upstream of the TPC before and after the SFC upgrade. The black line represents the beam intensity measured at the exit of the old SFC, and the red represents the beam intensity after the upgrade. Due to differences in the measured positions, the TOF spectrum of the old SFC is adjusted to align with that of the new SFC.} 
 \label{fig:SFC_compare}
\end{figure}

The PHITS calculation used the phase-space distribution at the exit of the polarized beam branch, whose absolute value was normalized by the flux at Pos.1.
The flux ratio decreased by 26\% from upstream M1 to downstream M2 but did not change thereafter. The decreases in M1 and M2 can be attributed to the divergence used in the calculation being smaller than the actual values.
The TOF spectra of the BM before and after the upgrade at Pos.4 are shown in Fig.~\ref{fig:SFC_compare}. The BM count increased by a factor of 3.2 times over that of the previous setup.
Since the BM counts are proportional to $1/v$ as well as the number of the neutron decay, this increase can be directly equated to an increase of the neutron decays.
Consequently, the rise timing in TOF shifted 20\% backward compared to the previous time owing to the alignment changes.

\subsection{Flipping Efficiency}
\label{sec:flip_eff}
\begin{figure}[ht]
\centering
  \includegraphics[width=13cm]{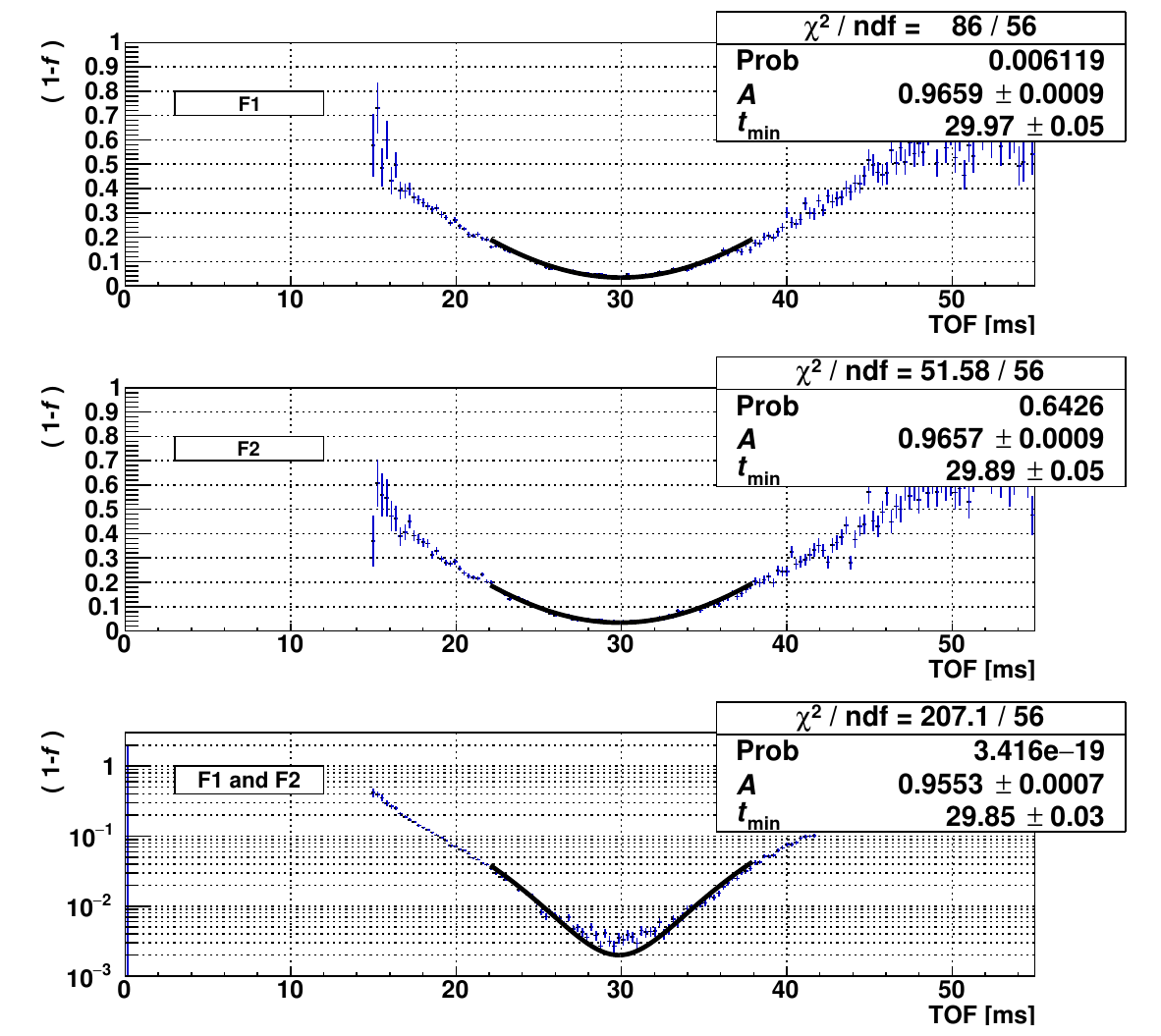} 
\caption{TOF spectra of the adjusted flipping rate $(1-f)$ fitted with the function of Equation~(\ref{eq:sine}). The figures respectively depict (Top) F1, (Middle) F2, and (Bottom) the operation of both F1 and F2 simultaneously.}
  \label{fig:F1Flip}
\end{figure}

Under condition of Pos.4 in Section~\ref{meas_flux}, we optimized the RF frequencies and currents of F1 and F2 to obtain a higher flipping efficiency. When the frequency satisfied Eq.~(\ref{eq:By}), $f$ 
became Eq.~(\ref{eq:sine}), and  the flipping efficiency exhibited a simple sine function on the TOF when the RF current was fixed. 
Thus, we obtained the spectra of $(1-f)$ on the TOF by considering the ratio of with the RF on to off and fitting them with the function 
\begin{equation}
1 - A \sin^{2} \left( \frac{t_{\mathrm{tof}}}{t_{\mathrm{min}}} \cdot\frac{\pi}{2} \right),
    \label{eq:fit}
\end{equation}
where $A$ denotes the flipper perfection. 
In this context, $t_{\rm tof}$ represents TOF of pulsed neutrons, whereas $t_{\rm min}$ denotes the moment when $(1-f)$ reaches its minimum, at which point $f$ equals $A$. Initially, we tuned the RF frequency to maximize $A$.
Following the frequency optimization, the RF current was adjusted to minimize the $(1-f)$ spectrum at a specific time of 30 ms (corresponding to a wavelength of 0.61 nm) so that 
the large part of the sine curve can be clearly observed in the TOF range utilised. The spectra for F1, F2, and F1 and F2 operating simultaneously are shown in Fig.~\ref{fig:F1Flip} from the top to bottom, respectively. The frequencies that produced the maximum $A$ were 32.2~kHz for F1 and 30.6~kHz for F2.
The spectra were fitted using Eq.~(\ref{eq:fit}) for F1 and F2, and square of Eq.~(\ref{eq:fit}) for the simultaneous operation of F1 and F2.
The maximum flipping efficiencies corresponding to $A$ in the fit were 96.6\% for both F1 and F2, and 95.5\% for the simultaneous operation. 
The minimum values for F1 and F2 were 3.96~$\pm$~0.18\% and 3.85~$\pm$~0.18\%, respectively, which are 16\% higher than $(1-A)$. The simultaneous operation worsened to 0.330~$\pm$~0.020\%, which is 63\% higher than the value indicated by $(1-A)^2$ and 2.2 times larger than the product of single operations. The depolarization of the polarizing mirror was suspected to be the cause. A quantitative discussion is provided in Section~\ref{sec:SFCdiscussion}.

\subsection{Neutron bunching operation}
\label{sec:bunching_operation}
\begin{figure}[ht]
 \centering
  \includegraphics[width=10cm]{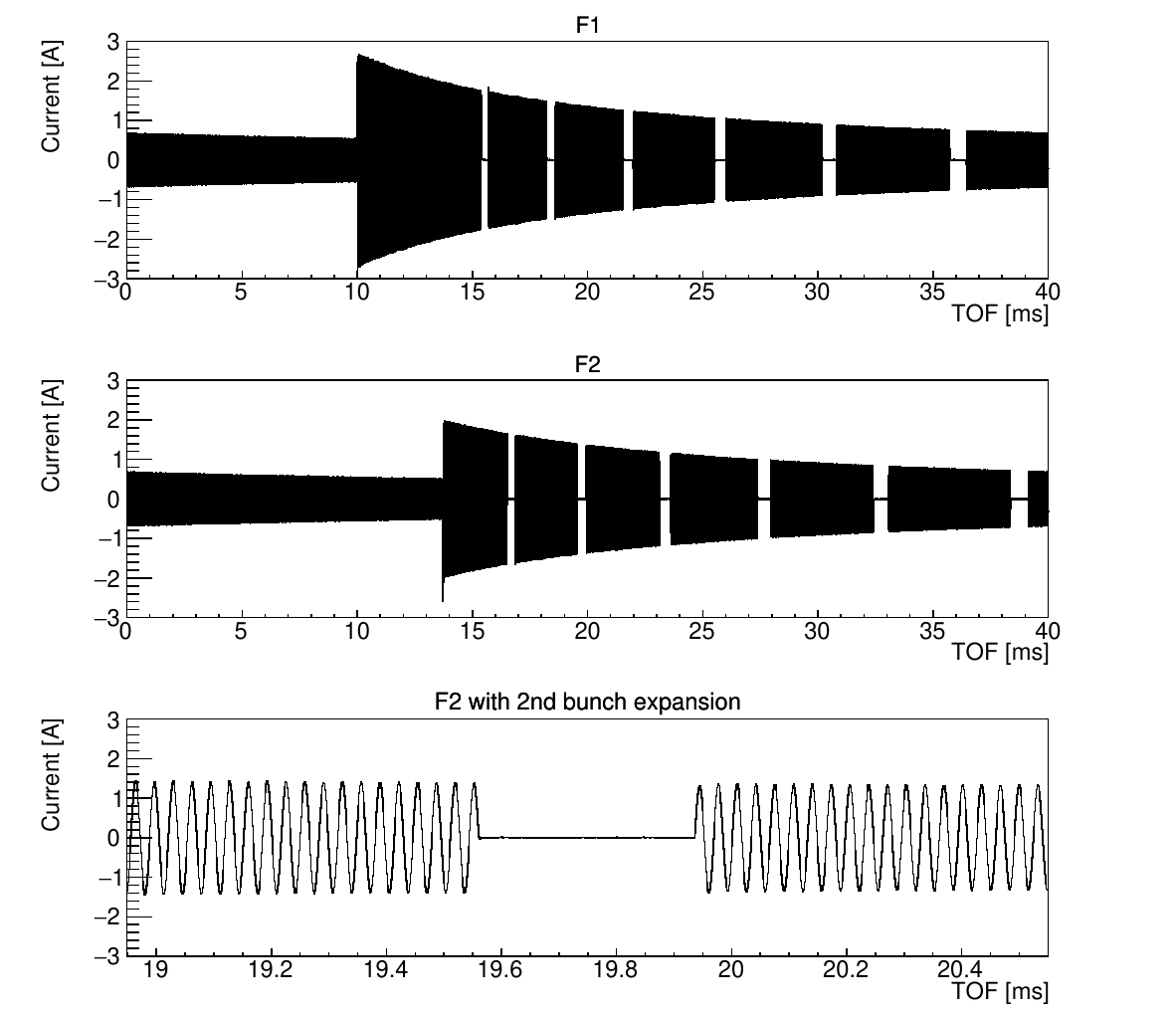} 
 \caption{Flipper current measured with a current meter. (Top) F1, (middle) F2, (bottom) an expanded view of the second bunch from F2.
 }
  \label{fig:Waveform}
\end{figure}

When dumping neutrons in all TOF regions, the RF current is inversely proportional to the neutron velocity; that is, $I(t_{\rm tof})=I_{0}/t_{\rm tof}$, where $I_{0}$ is the initial current at the time of pulse-frame switching.
The frame switching was set to occurred at 10~ms in F1 and was accordingly adjusted to 13.7~ms in F2.
When a bunch is created, the RF currents are deactivated.
The operation mode in which all beams enter the TPC by deactivating the RF is referred to as the passing mode. On the contrary, the dumping mode describes the condition in which all neutrons are absent during the application of RF currents. The specific condition used in the neutron lifetime experiment is termed the bunching mode. The RF currents monitored during the bunching mode are depicted in Fig.~\ref{fig:Waveform}.
The RF currents were controlled by the voltage applied to the amplifier. The phase between voltage and current was shifted by \qty{75}{\degree}, indicated in Table~\ref{table:RFcoils}. 
Taking this into account, voltage switching was performed when the current was zero. 

The detection length along $Z$-axis must be sufficiently small to measure the rise time of a bunch. Thus, instead of the BM, a two-dimensional detector (RPMT~\cite{hirota2005}) using a 0.4-mm thick ZnS scintillator for neutron detection was employed. The detection efficiency of RPMT was 
$\varepsilon_{\mathrm{RPMT}} = 1 - \exp\left( - 0.97\cdot\lambda~\mathrm{[nm]}\right)$.

The RPMT was installed at a position corresponding to the center of the TPC.
Because the RPMT could not receive all beams in terms of counting, a pinhole was installed at the SFC lead exit, and a scan was performed in 1~mm steps. 
This measurement enabled the three-dimensional reconstruction of the position and divergence distribution of the neutron beam from the SFC and its use in simulations of the neutron lifetime experiment.

\begin{figure}[ht]
 \centering
 \includegraphics[width=12cm]{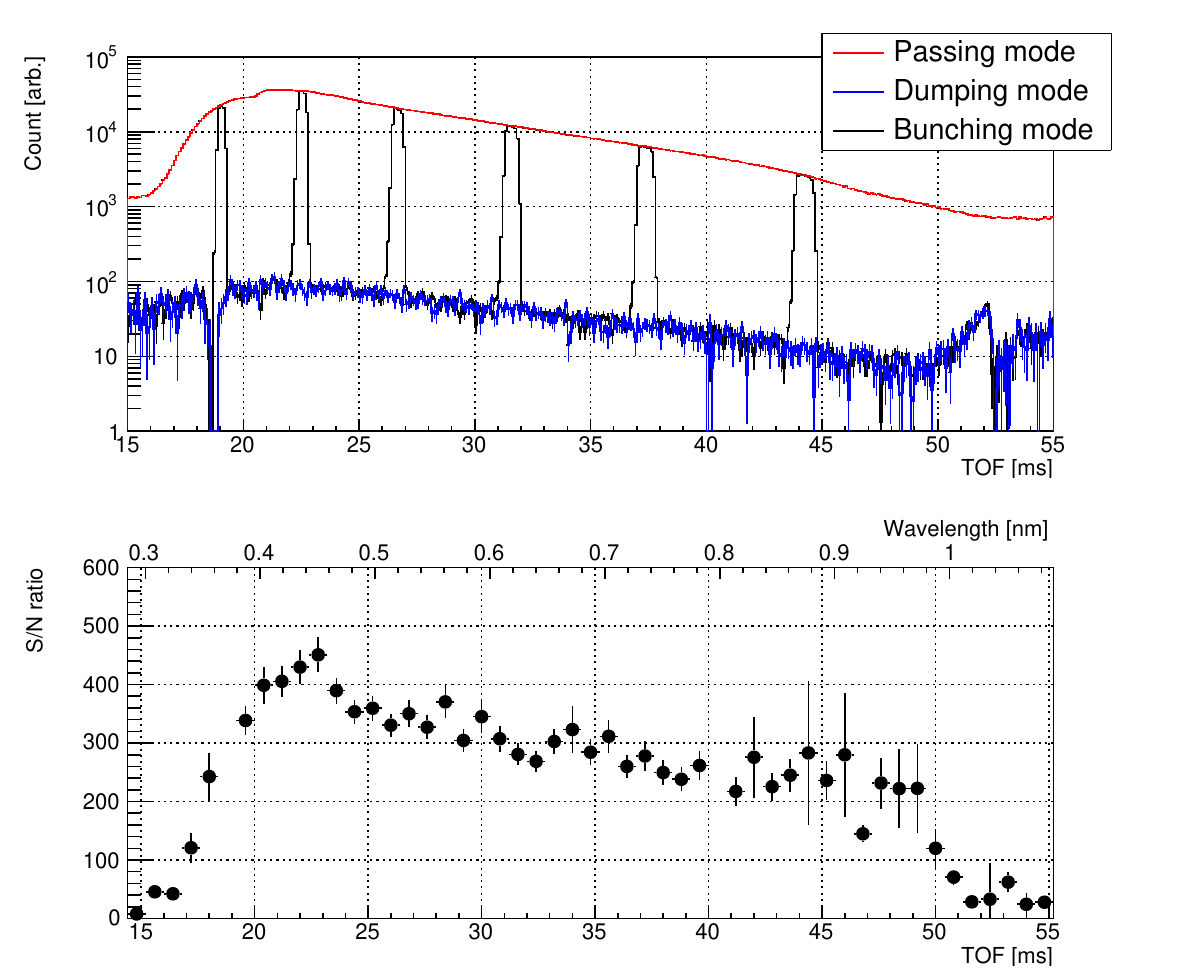}
 \caption{(Top) TOF spectra measured with RPMT. The colors correspond to Passing mode (red), Dumping mode (blue), and Bunching mode (black). (Bottom) The S/N of the SFC, representing the Passing mode divided by the Dumping mode}
 \label{fig:RPMT_TOF}
\end{figure}

Figure~\ref{fig:RPMT_TOF} shows the TOF distribution obtained using the RPMT in the passing, dumping, and bunching modes. For convenience, the TOF of 0--15~ms is displayed as 40--55~ms. In the bunching mode, because neutrons hitting the magnetic mirror or beam dump cause $\gamma$-rays to become the background source, the SFC bunch interval was adjusted to ensure that other bunches were not present at positions generating backgrounds while a certain neutron bunch was in the TPC. The lengths of the neutron bunches were adjusted to 40~cm and the distance between the end of one bunch and the tip of the next bunch was set to 3.4~m. In the present upgrade, the number of bunches in a pulse was increased from the previous five to six because longer-wavelength neutrons can now be used. 
The rise time of the bunch was obtained by fitting with an error function, and its 1$\sigma$ was \qty{63}{\micro\second} for the first peak and \qty{97}{\micro\second} for the sixth peak. These rise times approximately correspond to the time it takes for neutrons to travel through a distance of 60~mm, which is equivalent to twice the sigma ($2\sigma$) of the RF magnetic field width, as shown in Fig.~\ref{fig:Flipper}(D). The S/N of the SFC,
obtained as the ratio of the passing mode to the dumping mode, 
is shown at the bottom of Fig.~\ref{fig:RPMT_TOF}. The S/N values were in the range of 250--400 for all the bunch regions.

\subsection{Polarization Measurement}
\label{sec:polarization}
\begin{figure}[ht]
 \centering
  \includegraphics[width=10cm]{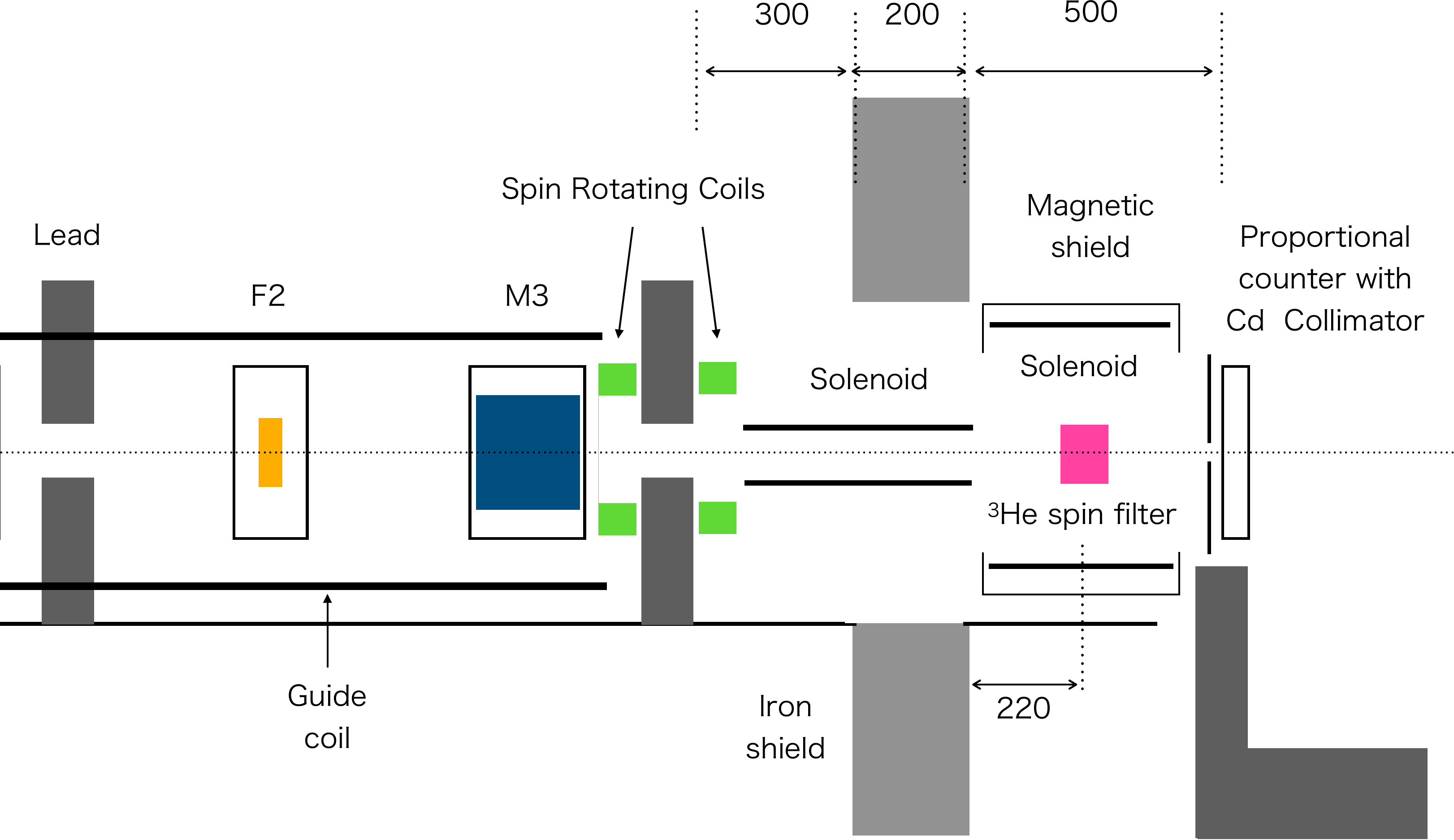}
 \caption{Setup for the polarization measurement (side view). Two spin rotation coils were installed before and after the lead at the downstream of the SFC. A solenoid coil was installed to connect the magnetic field downstream of it, and $^{3}$He spin filter was stored in the solenoid in the magnetic shield. Neutrons transmitted through the spin filter were counted by the $^3$He proportional counter.}
 \label{fig:polmeasurement_setup}
\end{figure}

Based on the discussion presented in Section~\ref{sec:SFCprinciple}, neutron polarization after passing through the SFC can be used to evaluate the performance of the magnetic mirror and flipper. Therefore, we measured the polarization at the front of the TPC by using a $^3$He spin filter. The experimental setup is shown in Fig.~\ref{fig:polmeasurement_setup}.

The $^3$He spin filter is a device that controls neutron polarization, utilizing the property that the neutron capture cross-section varies greatly depending on the neutron polarization direction. In this study, we used an $^3$He filled glass vessel with diameter and length of 40, and 90~mm, respectively, filled at a pressure of $3.1\times10^{5}$~Pa (HANABISHI, $\rho d = 25.8~\rm{atm \cdot cm}$)~\cite{okudaira2020}. 
The spin filter was placed in the center of a solenoid magnet covered with a magnetic shield, and a magnetic field of 1.5~mT was applied.
The neutrons were polarized along the $Y$-direction (vertical direction) using a guide coil. Two spin rotation coils with diameter of 100~mm were installed before and after the lead at the SFC exit to adiabatically rotate the polarization direction of the neutron beam from the $Y$-axis to the $Z$-axis. When the magnetic field of the spin rotation coils was applied at 3.6~mT, the depolarization for neutrons with a velocity of 1000~m/s (wavelength 0.4~nm) is estimated to be $2\times10^{-3}$, which is sufficient for the measurement precision in this study.
The neutron beam passing through $^3$He was detected using a $^3$He proportional counter collimated to a 10-mm square.

\begin{figure}[ht]
 \centering
 \includegraphics[width=10cm]{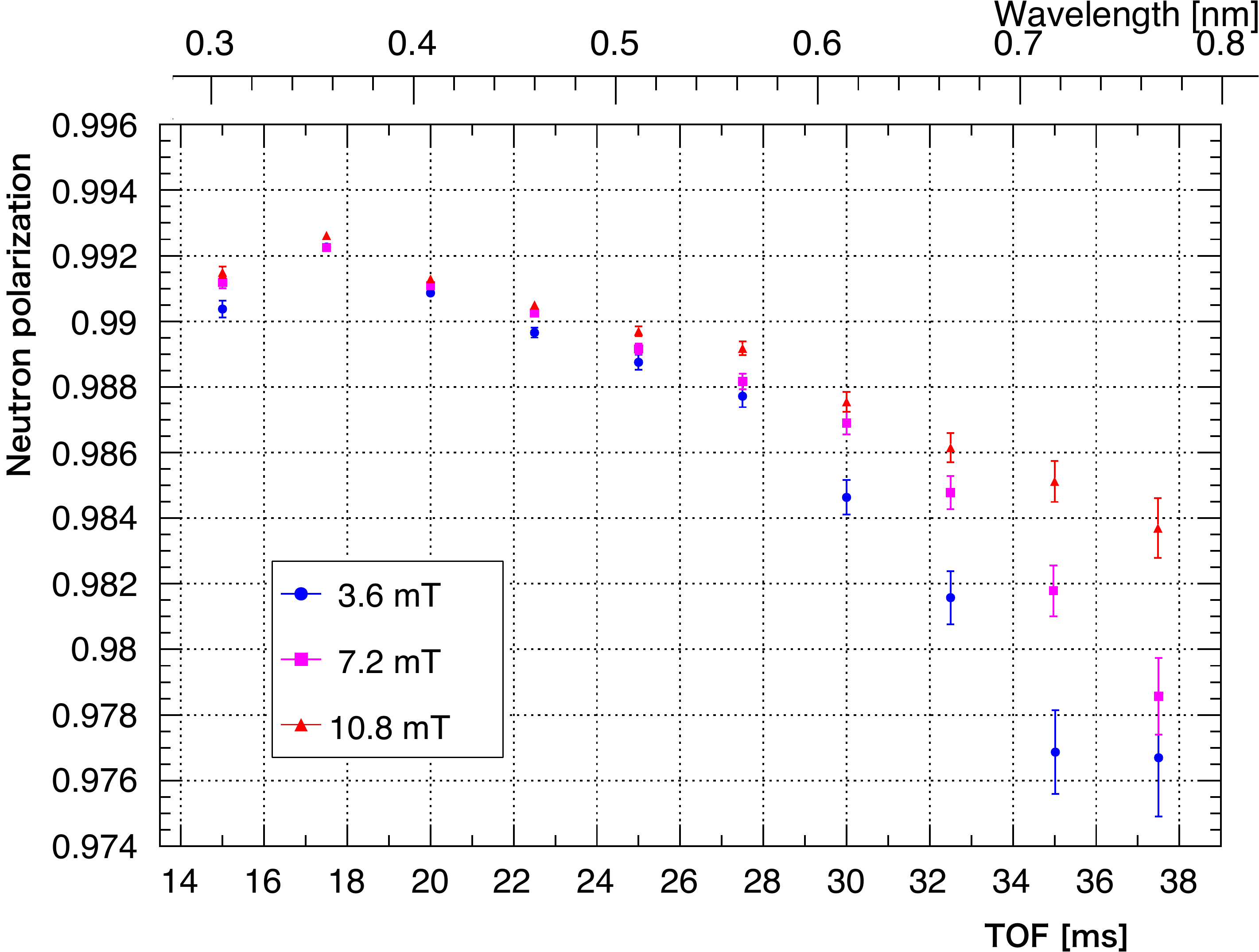}
 \caption{Polarization with the current of the spin rotation coils. In the figure, blue circles, pink squares, and red triangles correspond to the central magnetic fields of the spin rotation coil at 3.6, 7.2, and 10.8~mT, respectively.}
 \label{fig:neutron_polarization_drawsame}
\end{figure}

Helium-3 was polarized using the spin-exchange optical pumping (SEOP) method. 
The polarization of neutrons and $^3$He can be determined by measuring their transmission through $^3$He in states of polarized, spin-flipped through adiabatic fast passage, and finally unpolarized. The detail of this method is described in reference~\cite{okudaira2020}. The initial $^3$He polarization was approximately 70\% and decreased with time. The results of the neutron polarization measurements are shown in Fig.~\ref{fig:neutron_polarization_drawsame}. The measurements were performed while changing the magnetic field of the spin-rotation coils to 3.6, 7.2, and 10.8~mT. 
A polarization is 99.2 \% for the neutrons arriving at 18~ms. From there, the polarization tends to decrease with slower neutrons, and a magnetic field dependence of the rotating coil is observed for polarization with wavelengths longer than 0.5 nm. 
Because the effect of non-adiabatic depolarization should be less than $2\times10^{-3}$ and potentially even smaller for longer wavelengths or stronger magnetic fields, it is unlikely that depolarization during transit is the cause of the observed reduction.

\subsection{Discussion on SFC Performance}
\label{sec:SFCdiscussion}
In this subsection, we discuss the performance of the developed SFC, considering the observed results and identifying its limitations and potential improvements.

The S/N measured by RPMT, as shown in Fig.~\ref{fig:RPMT_TOF} was 250--400, while the ideal S/N of the SFC estimated in Section~\ref{sec:SFCprinciple} is $\sim$1000. The lower S/N can be attributed to the depolarization in the magnetic mirror, as inferred from the measurements of the spin flippers in Section~\ref{sec:flip_eff} and the polarization in Section~\ref{sec:polarization}. 
The polarization measured by the $^3$He spin filter was $P = 97.6 \text{--} 99.2\%$. As an example, at 30~ms, the value was 98.45~$\pm$~0.05\%, which is significantly worse compared to the estimation of the non-adiabatic depolarization of $2\times10^{-3}$ discussed in Subsection~\ref{sec:polarization}.
The polarization is degraded when a spin-flip reflection occurs at the magnetic mirror.
In the calculations in Section~\ref{sec:SFCprinciple}, we treated $R_{\rm +-}$ and $R_{\rm -+}$ as zero. 
If $R_{\rm +-}$ is assumed to be 0.5\%, the polarization after M3 is expected to be $P=98.9\%$.

\begin{figure}[ht]
\centering
  \includegraphics[width=10cm]{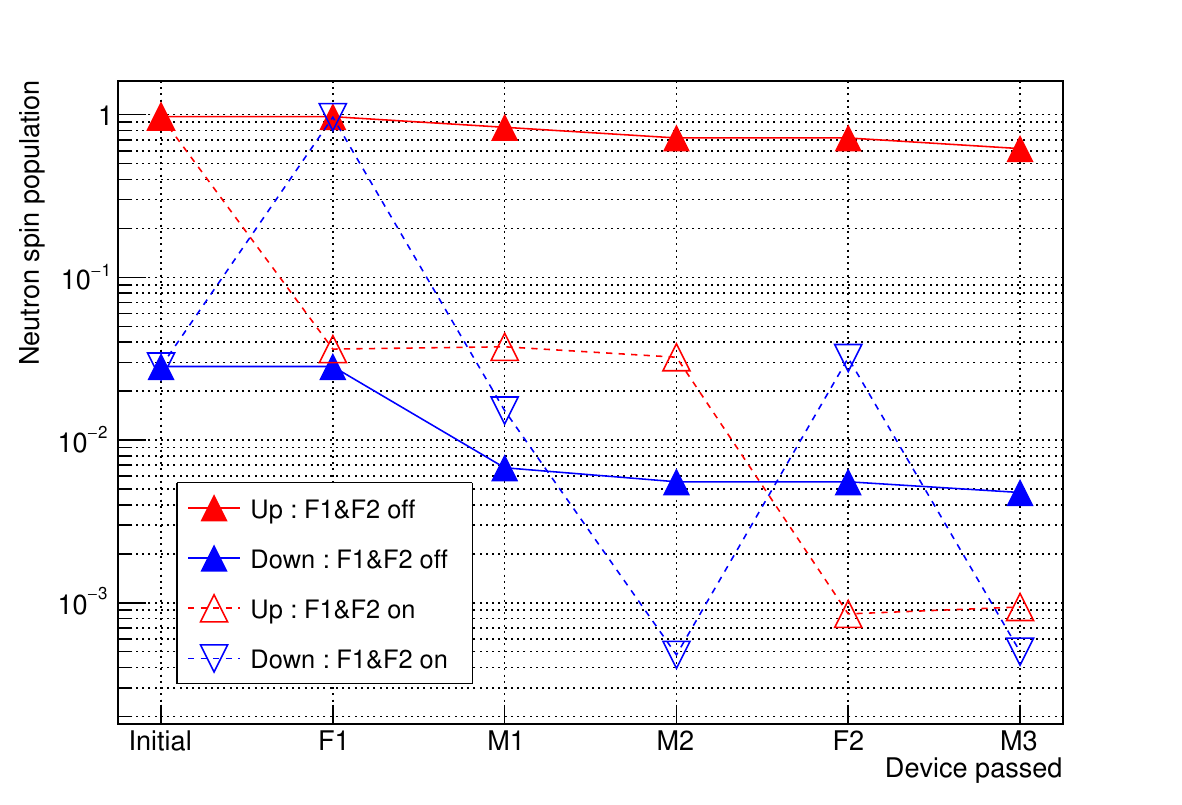}
 \caption{Neutron quantities after passing through the device, obtained with the parameters resulted from the global fit listed in Table~\ref{table:GlobalFit}. The upward (red) and downward (blue) triangles represent $n_{\rm +}$ and $n_{\rm -}$, respectively. The closed triangles with solid lines indicate without operating the flippers, while the open triangles with dashed line represents the case where both F1 and F2 are operated.}
  \label{fig:M1M2M3real}
\end{figure}

Here, we evaluate the flipping efficiencies and mirror reflectivities including the spin-flip reflection to explain the experimental values. For this analysis, we use a global fitting, which optimize parameters in a model function that minimize $\chi^2$, based on measured value and their associated errors.
Four parameters in our model function of Eq.~(\ref{eq:SFC}) are set as free parameters: $f_1$ and $f_2$ for the flipping efficiencies of F1 and F2, $R_{--}$ for the reflectivity of the magnetic mirror for spin $-$, and $R_{+-} = R_{-+} = R_\text{flip}$ for reflectivity with a spin flip. 
The optimal values of these parameters are derived through the global fit with four measured values: the lowest values for F1, F2, their simultaneous operation as illustrated in Fig.~\ref{fig:F1Flip}, and the polarization of $P=98.45~\pm~0.05\%$ measured with the $^3$He spin filter at 30~ms (0.61 nm) with a magnetic field of 3.6~mT in the spin rotation coil.
Since the flipper measurements were performed without spin rotation coils, we use the polarization value with the minimum coil field of 3.6~mT. 
The results are presented in Table~\ref{table:GlobalFit}, and the corresponding spin states are shown in Fig.~\ref{fig:M1M2M3real}.

\begin{table}[ht]
\centering
\caption{Results of the global fit}
\begin{tabular}{c|c} \hline 
$f_{1}$ & 99.30 ~$\pm$ ~0.17 \%\\ 
$f_{2}$ & 98.85 ~$\pm$  ~0.22 \% \\ 
$R_{\rm --}$ & 1.42 ~$\pm$  ~0.11 \% \\
$R_{\rm flip}$ & 0.65 ~$\pm$ ~0.02 \% \\ \hline
S/N & 466 \\ \hline
\end{tabular}
\label{table:GlobalFit}
\end{table}

\begin{figure}[!ht]
\centering
 \includegraphics[width=12cm]{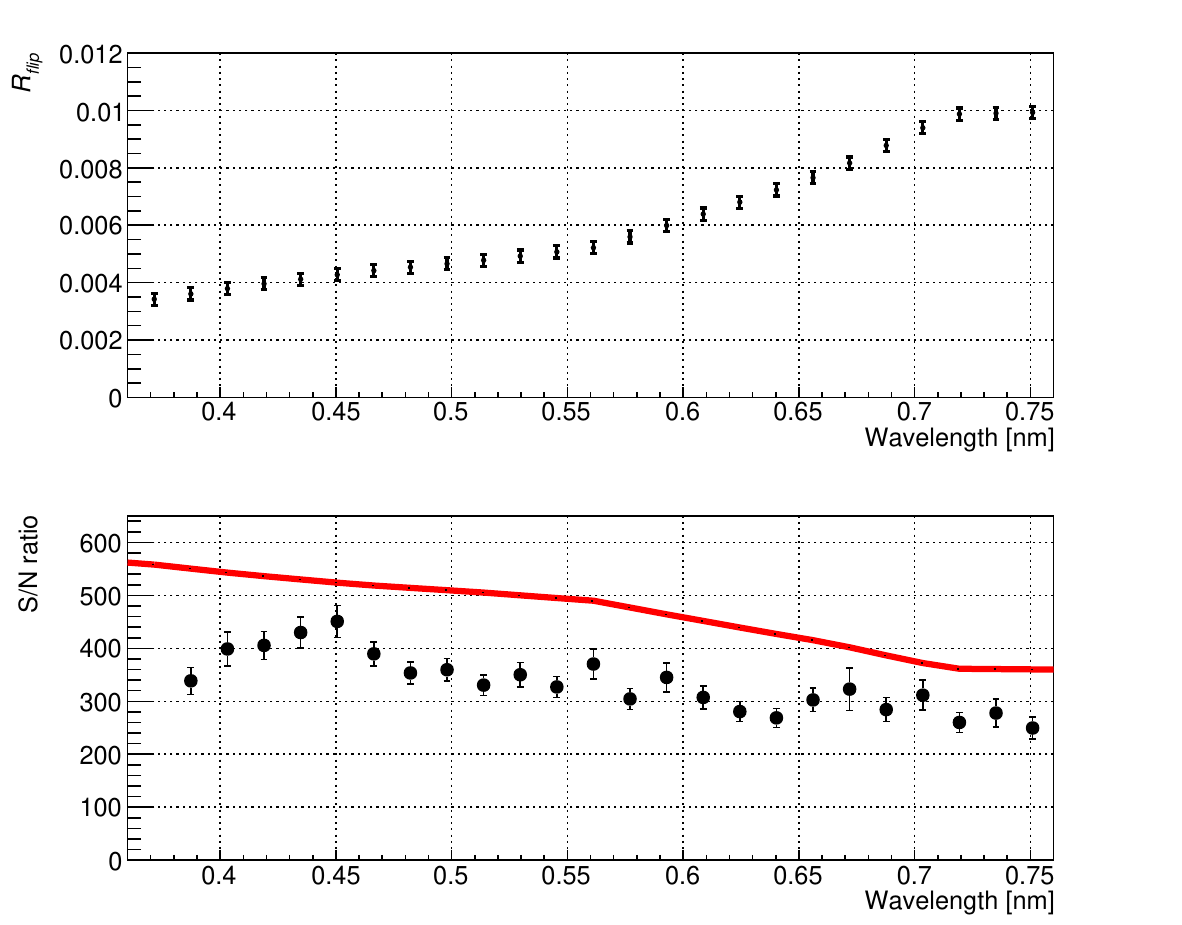}
 \caption{(Top) Wavelength dependence of $R_{\rm flip}$ obtained by the fit (Bottom) S/N derived from $R_{\rm flip}$ shown in the top (red line) with measured values (black circles)}
  \label{fig:Rflip_SN}
\end{figure}

The spin-flip efficiencies of $f_1$ and $f_2$ obtained by the fit are approximately 99\%, which is comparable to the previous estimation of the flipping efficiency in Section~\ref{sec:flipcalc}. The value of $R_{--}$ of 1.42\% appears to be reasonably consistent with the specification in Table~\ref{tableMirror}. Derived $R_{\rm flip}$ is 0.65~$\pm$~0.02\%, which is in closely aligns with the reported value of 0.56~$\pm$~0.03\% for Fe/Si~($m=3$) at 35~mT magnetic field in reference~\cite{klauser2016}. The S/N of 466 obtained from the fit is closer to the experimental value of 303 than the ideal value of 1140. 

The above analysis is for a certain wavelength of 0.61~nm. There appears to be the wavelength dependence of the S/N in Fig.~\ref{fig:RPMT_TOF} and the polarization in Fig.~\ref{fig:neutron_polarization_drawsame}.
To investigate them, we derive $R_{\text{flip}}$ for each wavelength by fitting the polarization of Fig.~\ref{fig:neutron_polarization_drawsame} with fixing the other parameters ($f_1$, $f_2$, and $R_{--}$). 
The results obtained are shown in the top of Fig.~\ref{fig:Rflip_SN}, where $R_{\text{flip}}$ shows a wavelength dependence that increases with wavelength. This is consistent with the trend of the momentum transfer dependence reported in reference~\cite{petoukhov2023}. From this wavelength dependence of $R_{\text{flip}}$, with the other fixed parameters, the TOF dependence of the S/N are derived. The results are shown as a red line in the bottom of Fig.~\ref{fig:Rflip_SN} with the measured S/N shown in Fig.~\ref{fig:RPMT_TOF}. 
While the S/N derived from Fig.~\ref{fig:Rflip_SN} based on polarization is somewhat higher than the measured values, it exhibits a similar decreasing trend. This suggests that the reduction in the S/N could be attributed to an increase in $R_{\text{flip}}$.

In conclusion, spin-flip reflection at the polarised mirrors limit the S/N and polarization of the SFC. 
According to reference~\cite{petoukhov2023}, a magnetic field of over 200~mT is required to reduce depolarization to below 0.1\%. With an appropriate magnetic field, it is expect that $R_{\rm flip}$ can be reduced to 0.1\%, which would achieve the S/N of 730 and polarization of 99.8\%. This enhancement would significantly improve the performance of the SFC compared to the current setup.

\section{Signals in TPC}
\label{sec:TPC}
\begin{figure}[ht]
\centering
  \includegraphics[width=10cm]{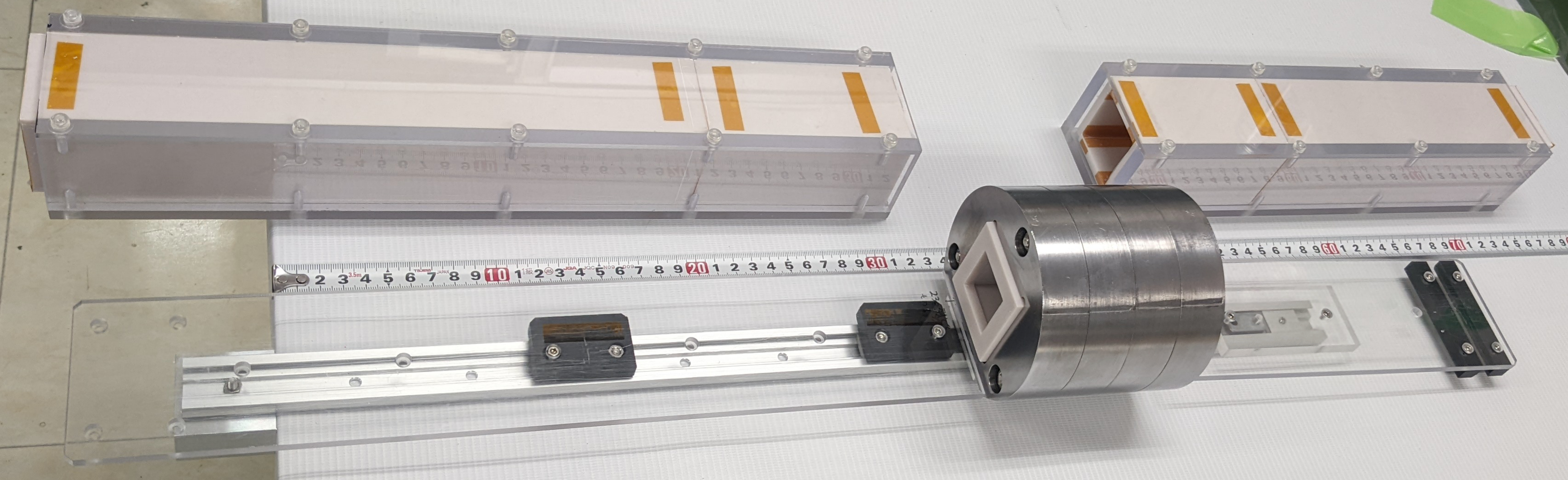} 
 \caption{Installed lead collimator between the LiF conduits upstream of the TPC. The left side of the photo is the upstream.}
  \label{fig:LeadLiF}
\end{figure}

In this Section, we discuss the neutron signals in the TPC. In the neutron lifetime experiment, 
the neutrons were shaped into bunches with a length of 40 cm at the SFC and transported to the TPC, which was housed in a vacuum vessel covered with lead shielding and cosmic ray veto counters.
The upgrade of the SFC resulted in a 3.2-fold increase in the beam quantity upstream of the TPC in terms of $\beta$ decay events. However, this also increased the neutron beam width and divergence, as well as the amount of $\gamma$-rays, owing to neutron capture by the magnetic mirror. To mitigate these increases, a new lead collimator, 100~mm in length and 36~mm in inner diameter, was installed between the guides upstream of the TPC made of the LiF tiles. 
The lead collimator was located 300 to 400~mm upstream from the TPC. LiF tiles with a 30~mm square hole were installed upstream and downstream of the lead collimator to define the size of the neutron beam entering the TPC.
The inner surface of the lead collimator was also covered with LiF to absorb the scattered neutrons.
The photo of the collimator and LiF guides are shown in Figure~\ref{fig:LeadLiF}.
 
\begin{figure}[ht]
\centering
  \includegraphics[width=10cm]{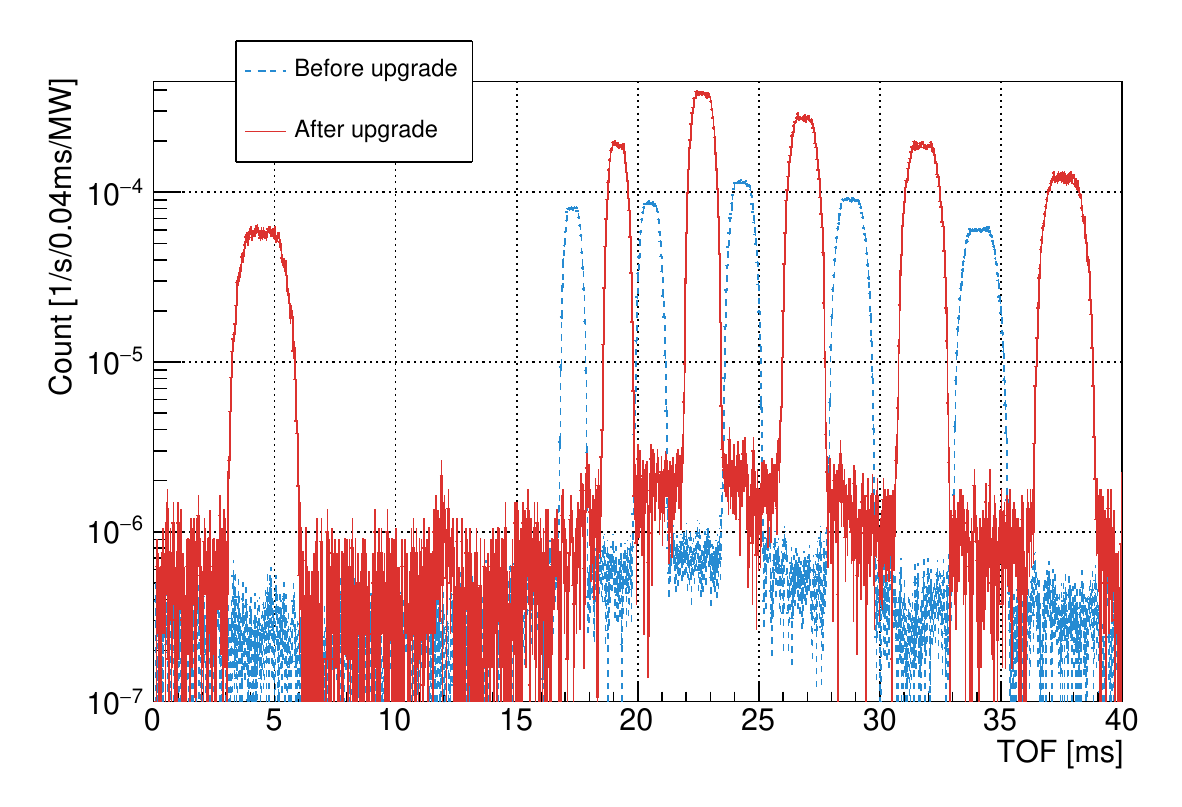} 
   \caption{TOF spectra of the bunched neutron beam measured by the TPC, selecting the $^3$He(n,p)$^3$H reaction. The blue dotted line represents the spectra before the SFC modification, while the red solid line represents the spectra after the modification. The beam power was normalized to 1~MW and the $^3$He number density to $1.5\times10^{19}$ atoms/m$^{3}$.
 }
  \label{fig:TOFplotHe}
\end{figure}

The impact of the SFC upgrade was clearly demonstrated in the TOF spectra of the $^3$He(n,p)$^3$H reaction with the TPC both before and after the upgrade, as shown in Fig.~\ref{fig:TOFplotHe}. The gas mixture was set to 83 kPa for He and 15 kPa for CO$_2$, which are the standard measurement settings for the lifetime measurements~\cite{hirota2020neutron}. In addition, $^3$He was introduced into the gas mixture, resulting in a number density of $\rho = 1.5 \times 10^{19}~{\rm atoms/m}^{3}$.
The upgrade resulted in a 3.2-fold increase in the $^3$He(n,p)$^3$H reactions as well, while maintaining comparable S/N and time widths, as before the upgrade ~\cite{ichikawa2022}. 
A small peak emerged around 12~ms, accounting for about 1/1000 of the neutron bunches due to frame overlap. Although this region was utilized as a background in the lifetime analysis, it had no impact on the lifetime results.

\begin{figure}[ht]
\centering
  \includegraphics[width=10cm]{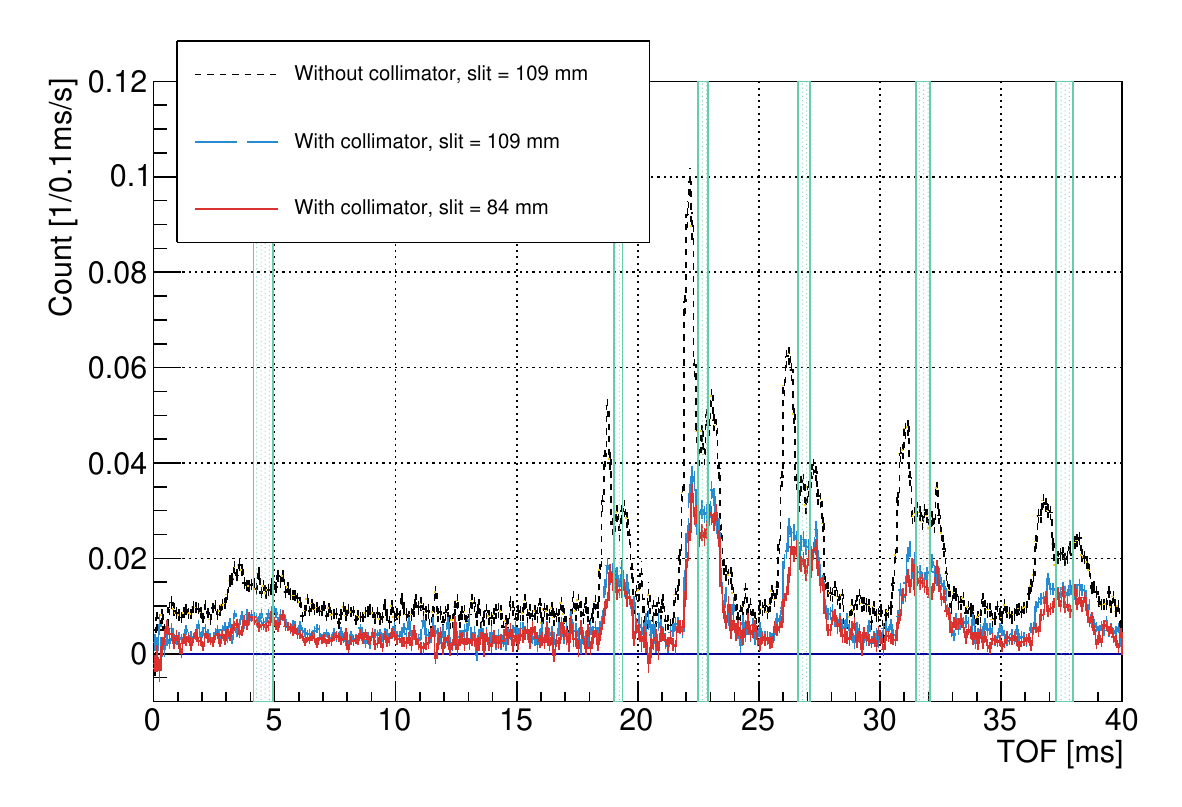} 
 \caption{TOF spectra of the $\beta$ decay candidate events. The black dashed line represents the spectrum without the lead collimator, the blue and red lines represents the spectrum with collimator, where the blue was taken with the 12-m slit vertical aperture of 109~mm, and the red was with 84~mm. The green bands indicate the fiducial areas where the neutron bunch passes through the TPC.
 }
  \label{fig:TOFplotBeta}
\end{figure}

The TOF spectra for selecting the $\beta$ decay candidate events using the new SFC are shown in Fig.~\ref{fig:TOFplotBeta}. 
As in the neutron lifetime experiment~\cite{hirota2020neutron}, the $\beta$ decay candidate signals were derived by subtracting the shutter closed signal from the shutter open signal.
Owing to the enlarged aperture of the SFC, peaks occurred in TOF corresponding to before and after the neutron bunch. This is considered to be caused by the neutrons hitting the LiF guides owing to the increased divergence. Installing a lead collimator mitigated these peaks. Additionally, events in the 10--15~ms range, where no neutrons were present, also decreased. 
These events are likely residual radiation from $^{20}$F and $^{8}$Li absorbed by the LiF guide because the shutters opened and closed every 1000~s and the causal radioactivities must decay faster than that time~\cite{hirota2020neutron}.

Because the divergence of the neutron beam is larger in the vertical direction than in the horizontal direction, reducing the vertical width from 109 to 84~mm using a slit located 12~m from the neutron moderator in the polarized branch further decreased the peaks before and after the bunches. By setting the vertical slit to 84~mm, the count of $^3$He(n,p)$^3$H events decreased by 10.1\% compared with the 109~mm condition. Consequently, under these settings, the neutron count increased by 2.8 times compared to that of the original SFC.

\begin{figure}[ht]
\centering
  \includegraphics[width=10cm]{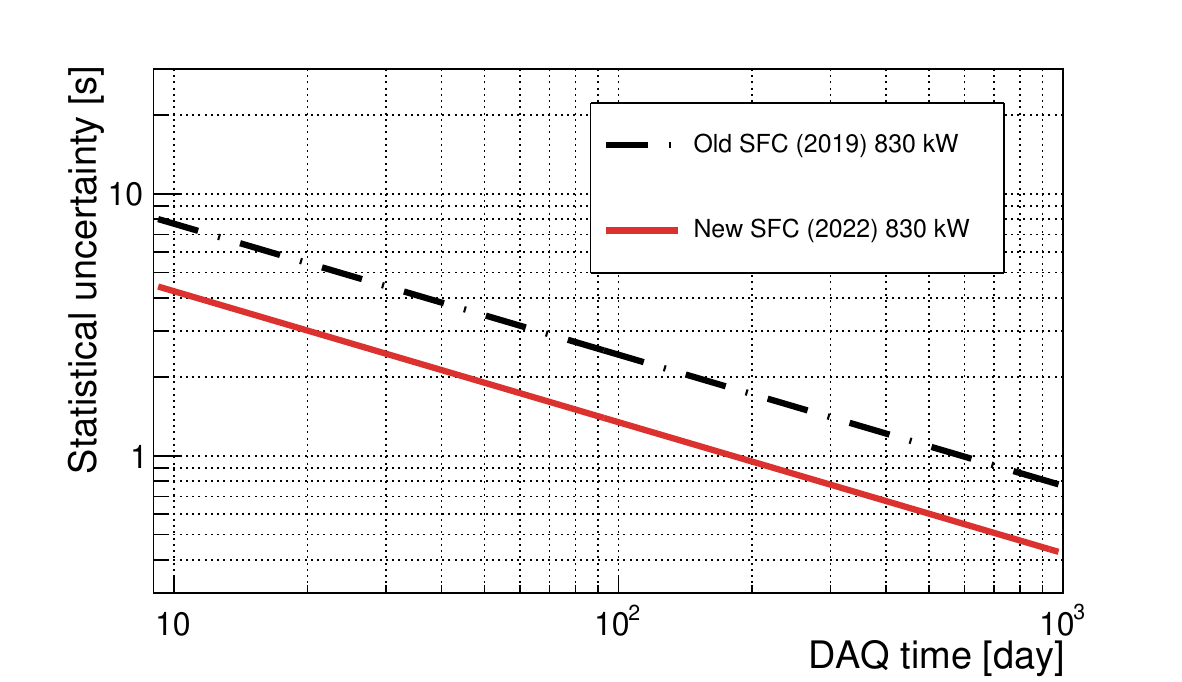} 
 \caption{Estimation of measurement time and sensitivity reached.
 }
  \label{fig:Stat}
\end{figure}

The statistical uncertainties under the optimal conditions (red in Fig.~\ref{fig:TOFplotBeta}) before and after the SFC upgrade, estimated from the number of $\beta$ decay candidate signals and backgrounds, are shown in Fig.~\ref{fig:Stat}. The beam power assumed in the calculation is 830~kW. 
The measurement time required to achieve a statistical uncertainty of 1~s is reduced from 590~days to 170~days due to the SFC upgrade. Note that these durations do not account for calibrations or additional procedures performed during the physics experiments.

\section{Summary}
\label{Summary}
In the neutron lifetime experiment using the pulsed neutron beam at J-PARC, the statistics was limited by the acceptable beam size at the SFC. In this study, the sizes of the magnetic mirrors and spin flippers were enlarged to improve the neutron intensity. The magnetic container was made of neodymium and iron magnets, ensuring that the magnetic field was greater than 46~mT over the entire region. The neutron flux before TPC measured by the BM was converted to 1~MW at $1.4\times10^6$~cm$^{-2}$/s, which is 75\% of the value calculated by PHITS. 

The structure of the spin flipper was redesigned to mitigate the effects of flinging fields from the magnetic container. A $B_0$-magnetic field was generated inside the ferrite magnets in the iron shield box, and the RF coil was housed inside it. 
The timing of the neutron spin flip can be freely selected by amplifying the output of an arbitrary waveform generator to control the RF coil current.
The rise time in the bunching mode was \qtyrange{63}{97}{\micro\second} at 1$\sigma$ of the error function, which is equivalent to twice the standard deviation of the magnetic field distribution by the RF coil.

The neutron reduction rate under optimized RF conditions was approximately 3.9\% for both F1 and F2, and 0.33\% for simultaneous operation. The result of the simultaneous operation was 2.2 times worse than that expected from the square of the single-flipper operation. The spin polarization of the neutrons after SFC was measured using a $^{3}$He spin filter and obtained as 97.6--99.2\%. Further, wavelength dependence was observed, with slower neutrons exhibiting worse polarization. The polarization tended to increase as the magnetic field strength of the coil used for the spin rotation increased. 
From the measured values at the neutron wavelength of 0.61~nm, the flipper efficiencies ($f_1$ and $f_2$), reflectivity for spin $-$ ($R_{--}$), 
and spin-flip reflectivity ($R_{\text{flip}}$) were obtained via global fitting. The derived values for $f_1$ and $f_2$ were 99\% and 1.4\% for $R_{--}$, which are reasonable based on the design.
The obtained $R_{\text{flip}}$ was 0.65~$\pm$~0.02\%. It is considered that the magnetic field on the mirrors was not sufficient to have a sufficiently small $R_{\text{flip}}$ and could be improved by increasing the magnetic field of the mirror container.

The $^3$He(n,p)$^3$H reaction count in the TPC, normalized by the power of the proton beam and the density of $^3$He numbers, was enhanced 2.8 times by improving the SFC. Consequently, the time to reach a statistical uncertainty of 1~s for the neutron lifetime experiment improved from 590 to 170~days with a proton beam power of 830~kW. This improvement in statistic will also contribute to the reduction of systematic uncertainties, such as background evaluation, fostering further advancements in the neutron lifetime experiments at J-PARC.

\section*{ACKNOWLEDGE}
This work was supported by JSPS KAKENHI Grant Number (16H02194, 19H00690, and 22H00140).
The neutron experiment at the Materials and Life Science Experimental Facility of J-PARC was performed under a user program (Proposal Nos. 2015A0254, 2019B0341, 2020A0223, 2021B0287 and 2022A0117) and an S-type project of KEK (Proposal Nos. 2014S03 and 2019S03).
Magnetic mirrors and flippers were installed at the CN3 beam port of KUR with the
approval of the Institute for Integrated Radiation and Nuclear Science, Kyoto University (Proposal Nos. 30086 and 31030).

\bibliography{Ref}
\end{document}